\documentclass[pdflatex,iicol]{sn-jnl}% Math and Physical Sciences Numbered Reference Style
\usepackage[numbers,sort&compress]{natbib}
\usepackage{graphicx}%
\usepackage[caption=false]{subfig}
\usepackage{multirow}%
\usepackage{amsmath,amssymb,amsfonts}%
\usepackage{amsthm}%
\usepackage{newtxtext,newtxmath}
\usepackage[title]{appendix}%
\usepackage{xcolor}%
\usepackage{textcomp}%
\usepackage{manyfoot}%
\usepackage{booktabs}%
\usepackage{algorithm}%
\usepackage{algorithmicx}%
\usepackage{algpseudocode}%
\usepackage{listings}%
\providecommand{\burl}[1]{{\urlstyle{sf}\url{#1}}}

\let\orcidlogo\undefined
\usepackage{orcidlink}

\theoremstyle{thmstyleone}%
\theoremstyle{thmstyletwo}%

\theoremstyle{thmstylethree}%

\begin{document}

\title[HERO-D3QN for Inter-Cluster Routing]{HERO-D3QN: Inter-Cluster Routing with Energy-Aware Soft Constraints in Wireless Sensor Networks}


\author[1]{Weihao~Wang$^{\orcidlink{0009-0000-9225-841X}}$}\email{wwhwwh@smail.sut.edu.cn}

\author*[1]{Heng~Yang$^{\orcidlink{0000-0003-0417-6313}}$}\email{h.yang@sut.edu.cn}

\author[2]{Guangbo~Chen$^{\orcidlink{0009-0002-7773-8655}}$}\email{c183266140@gmail.com}

\author*[3]{Shanshan~Li$^{\orcidlink{0000-0001-8563-7198}}$}\email{sgdlss@sut.edu.cn}

\author[1]{Jiulin~Guo$^{\orcidlink{0009-0001-3908-8834}}$}\email{j.guo@smail.sut.edu.cn}

\author[4]{Ajay~Kaushik$^{\orcidlink{0000-0001-7163-2337}}$}\email{a.kaushik@derby.ac.uk}

\author[1]{Fangwei~Li$^{\orcidlink{0009-0004-6051-0405}}$}\email{f.li@smail.sut.edu.cn}

\author[1]{Junrui~Hao$^{\orcidlink{0009-0002-3672-2788}}$}\email{david\_hao@smail.sut.edu.cn}

\author[1]{Sixue~Liu$^{\orcidlink{0009-0002-2945-9096}}$}\email{lsx@smail.sut.edu.cn}

\affil*[1]{
	\orgdiv{School of Information Science and Engineering},
	\orgname{Shenyang University of Technology},
	\orgaddress{
%		\street{Street},
		\city{Shenyang},
		\postcode{110870},
%		\state{Liaoning},
		\country{China}
	}
}

\affil[2]{
	\orgdiv{Guangzhou Institute of Technology},
	\orgname{Xidian University},
	\orgaddress{
%		\street{Street},
		\city{Guangzhou},
		\postcode{510555},
%		\state{Liaoning},
		\country{China}
	}
}

\affil*[3]{
	\orgdiv{School of Mechanical Engineering},
	\orgname{Shenyang University of Technology},
	\orgaddress{
%		\street{Street},
		\city{Shenyang},
		\postcode{110870},
%		\state{Liaoning},
		\country{China}
	}
}

\affil[4]{
	\orgdiv{School of computing},
	\orgname{University of Derby},
	\orgaddress{
%		\street{Street},
		\city{Derby},
		\postcode{DE22 1GB},
%		\state{},
		\country{UK}
	}
}


\abstract{Inter-cluster forwarding in wireless sensor networks can concentrate energy consumption on frequently selected relays, while hard energy thresholds restrict forwarding choices as residual energy declines. We propose Hierarchical Energy-balanced Routing Optimization based on a dueling double deep Q-network (D3QN), termed HERO-D3QN. Given the cluster structure, the method combines direction- and cost-based relay screening with energy-aware soft constraints. Relays below an adaptive energy threshold remain eligible if the other selection criteria are met. An enhanced state representation captures relay energy, progress toward the sink, potential forwarding load, and hotspot risk, and a multi-objective reward guides learning. In the comparison with D3QN using the same node locations, HERO-D3QN delayed first-node death (FND) from time step 629 to 1269 and reduced peak residual energy variance by 55.5\%. In the ablation study, the full method achieved the latest FND and half-node death (HND) and the largest cumulative transmitted volume (170.52 Mbit) among the tested variants. Across five seed settings in which node locations and the random processes used in training and evaluation varied jointly, the coefficients of variation of FND, HND, and transmitted volume were 8.26\%, 2.75\%, and 3.88\%, respectively. These results demonstrate the effectiveness of HERO-D3QN in delaying early node failure and reducing peak energy dispersion.}




\maketitle

\section{Introduction}\label{sec:introduction}

Wireless sensor networks (WSNs) collect and relay measurements for Internet of Things applications such as intelligent transportation, health monitoring, and smart cities. Sensor nodes typically operate with limited battery capacity, communication range, and processing resources. Recharging or replacing deployed nodes can be difficult, particularly in remote or harsh environments. Routing decisions therefore affect both the energy spent on communication and how long the network can remain operational.

Clustering and multihop forwarding address different aspects of this problem. The Low-Energy Adaptive Clustering Hierarchy (LEACH) rotates the cluster-head role to distribute energy expenditure across nodes~\cite{heinzelman2000leach,heinzelman2002application}. Hybrid clustering, hierarchical routing, and metaheuristic cluster-head selection extend these approaches to improve the distribution of energy use~\cite{panchal2021eehchr,zhou2024puag,yang2024mssomst}. Multihop forwarding can reduce the cost of long transmissions, but it also places additional demands on the nodes that relay the data. As residual energy and relay availability change, repeatedly selecting the same cluster heads can concentrate the forwarding load and accelerate their depletion~\cite{thakur2025aireview,lonkar2026hotspotreview}.

Reinforcement learning offers a way to adapt routing decisions to changing network conditions~\cite{sutton2018reinforcement,watkins1992qlearning,gaidhani2024mlreview}. Deep Q-networks (DQNs) estimate action values from high-dimensional observations~\cite{mnih2015humanlevel}, and learning-based routing has been studied in software-defined and distributed WSNs~\cite{wang2023cdrL,liu2023distributedneural}. Double Q-learning separates action selection from target evaluation to reduce overestimation~\cite{hasselt2016double}. A dueling architecture estimates state value and action advantages separately~\cite{wang2016dueling}. Graph representations, mean-field cooperation, and federated training provide different ways to incorporate topology information or coordinate learning across nodes~\cite{huang2022graphdrl,ren2024mefi,suresh2024federated}. WOAD3QN-RP uses whale optimization for cluster-head selection and a dueling double deep Q-network (D3QN) for multihop routing~\cite{yang2024woad3qn}.

In a WSN with a static sink, a cluster head choosing its next hop must account for progress toward the sink, the energy cost of forwarding, and the burden placed on the selected relay. A relay can offer a short transmission distance while having little energy left to forward subsequent packets. Excluding all low-energy relays can also reduce the available forwarding options as the network depletes. Given the cluster structure in each time step, we address this problem using Hierarchical Energy-balanced Routing Optimization based on D3QN (HERO-D3QN). The main contributions are as follows:

\begin{itemize}
    \item We develop a tiered next-hop selection rule with energy-aware soft constraints. The rule screens relays by direction and prioritizes them by progress toward the sink and the energy cost of a two-hop path relative to direct transmission. Soft penalties adjust action values according to residual energy relative to an adaptive threshold and the excess cost allowed by relaxed criteria. Relays with low residual energy remain eligible when the other selection criteria are met.
    \item We construct an enhanced state representation and a multi-objective reward for learning the action values used by the selection rule. The state describes residual energy, progress toward the sink, relative path length, hotspot risk, and indicators of potential forwarding load and onward forwarding opportunities. The reward combines node survival and transmitted volume with communication energy efficiency, energy balance, and relay risk; several reward weights vary with communication progress within each episode.
    \item We compare HERO-D3QN with a LEACH-based baseline, DQN, and D3QN using the same node locations, and examine variants of the action selector, state features, hotspot reward term, and intra-cluster forwarding. The analysis distinguishes early node loss from the persistence of the last surviving nodes and interprets energy dispersion alongside cumulative transmitted volume. We further characterize variation across five seed settings that jointly vary node locations and the random processes used in training and evaluation.
\end{itemize}

Section~\ref{sec:related-work} reviews energy-aware and learning-based routing for WSNs. Section~\ref{sec:system-model} presents the network and energy models and formulates the inter-cluster routing problem. Section~\ref{sec:method} describes how the given cluster structure is used in action-value estimation, relay selection, and replay-based learning. Section~\ref{sec:experiments} presents the comparative, ablation, and seed analyses. Section~\ref{sec:conclusion} summarizes the findings and outlines future directions.

\section{Related Work}\label{sec:related-work}

Cluster formation and packet forwarding both influence how energy expenditure is distributed across a WSN. LEACH rotates the cluster-head role so that different nodes take turns aggregating data and transmitting it to the sink~\cite{heinzelman2000leach,heinzelman2002application}. In classical LEACH, the cluster-head election threshold for node $n$ in time step $t$ is
\begin{equation}
T_t(n)=
\begin{cases}
\dfrac{p_{\mathrm{ch}}}{1-p_{\mathrm{ch}}\left(t\bmod \dfrac{1}{p_{\mathrm{ch}}}\right)}, & n\in\mathcal G_t^{\mathrm{CH}},\\
0, & n\notin\mathcal G_t^{\mathrm{CH}},
\end{cases}
\label{eq:leach-threshold}
\end{equation}
where $p_{\mathrm{ch}}$ is the desired fraction of cluster heads and $\mathcal G_t^{\mathrm{CH}}$ is the set of nodes that have not yet served as cluster heads in the current rotation cycle. This rule distributes the cluster-head role over successive time steps. In basic LEACH, each cluster head sends its aggregated data directly to the sink.

Subsequent work has explored hybrid clustering with hierarchical routing~\cite{panchal2021eehchr}, centralized clustering with multihop forwarding between cluster heads~\cite{tewelgne2025intercluster}, and metaheuristic clustering combined with minimum spanning trees~\cite{yang2024mssomst}. LPLL-LEACH revises cluster formation and inter-cluster forwarding to address energy consumption and delay~\cite{sun2023lpll}. ICMH-CHR sets thresholds for reclustering and cluster-head rotation based on the mean residual energy within a cluster~\cite{zeng2025icmh}. The energy-aware unequal clustering algorithm EAUCA forms smaller clusters near the sink and selects relays with low node degree to address the concentration of forwarding load near the sink~\cite{chauhan2021eauca}.

Radio energy models relate the cost of a transmission to the distance between nodes. Paulraj and Deepa~\cite{paulraj2024neurofuzzy} use a radio-link energy model in neuro-fuzzy routing, and Ambareesh et al.~\cite{ambareesh2025secure} consider energy consumption in a secure routing method based on type-2 fuzzy logic. Section~\ref{subsec:energy-model} specifies the first-order radio model used in this study. The energy cost of a hop is one consideration in relay selection, alongside the relay's residual energy and the load it may carry if several cluster heads use it.

Metaheuristic and graph-based methods also use spatial and topological information to organize forwarding. Zhou et al.~\cite{zhou2024puag} combine regional partitioning, unequal competition radii, and relay selection in a gray wolf optimization approach to energy balancing. In HRP-EE, Nguyen and Tan~\cite{nguyen2024hrpee} use Voronoi diagrams to form clusters, minimum spanning trees to construct intra-cluster communication paths, and Dijkstra's algorithm to determine routes between cluster heads and the sink.

Reinforcement learning adapts forwarding decisions using the rewards observed as network conditions evolve~\cite{sutton2018reinforcement,abadi2022rlsurvey}. In standard Q-learning~\cite{watkins1992qlearning}, the action-value update is
\begin{equation}
\begin{gathered}
Q(\mathbf{s},a)\leftarrow Q(\mathbf{s},a)+\alpha\bigl[r\\
+\gamma\max_{a'\in\mathcal A}Q(\mathbf{s}',a')-Q(\mathbf{s},a)\bigr].
\end{gathered}
\label{eq:q-learning}
\end{equation}
where $(\mathbf{s},a,r,\mathbf{s}')$ denotes a transition sample comprising the current state, action, reward, and successor state, respectively; $\mathcal A$ is the action space, $\alpha$ is the learning step size, and $\gamma$ is the discount factor. Applications include learning to select parent nodes in tree routing~\cite{kim2023enhancedtree}, accounting for energy consumption and quality of service in EEERP-RL~\cite{jaber2025eeerprl}, and using residual energy and packet loss in clustered routing under noisy conditions~\cite{chilukuri2025throughput}.

Deep Q-networks approximate action values from richer observations, with double Q-learning and dueling architectures providing different ways to estimate those values~\cite{mnih2015humanlevel,hasselt2016double,wang2016dueling}. In cellular handover, Wang et al.~\cite{wang2025marhandover} combine a multivariate DQN with a memory anchor repository to adapt decisions to the user equipment's context. For WSNs, WOAD3QN-RP combines whale optimization for cluster-head selection with learned multihop forwarding~\cite{yang2024woad3qn}. MeFi uses mean-field reinforcement learning to coordinate forwarding decisions~\cite{ren2024mefi}, while 3R applies multi-agent reinforcement learning to reliable routing in medical WSNs~\cite{hajar2023threeR}. Graph-based reinforcement learning represents network connectivity explicitly when making routing decisions~\cite{huang2022graphdrl}.

The network architecture also shapes the routing problem. In software-defined WSNs, a controller can use network observations to guide forwarding~\cite{wang2023cdrL,huang2022graphdrl,godfrey2023rlsdwsn}. With a mobile sink, RPP-RNN uses hotspot information to plan the sink trajectory~\cite{saravanan2025rpp}, while Al-Jerew et al.~\cite{aljerew2023mobile} use Q-learning to select cluster heads for data gathering while limiting relay hops and accounting for the sink's tour length. Federated approaches coordinate the training of routing models across participating nodes~\cite{suresh2024federated,sattibabu2025federated}.

These studies address energy-aware routing through different combinations of clustering, forwarding, and network control. This study focuses on the forwarding decision made after clusters have formed. Given a static sink and the cluster structure for the current time step, a cluster head chooses a reachable relay or transmits directly to the sink. HERO-D3QN uses candidate features to represent progress toward the sink, relative path length, residual energy, and hotspot risk. The reward weights vary with communication progress within each episode, and soft penalties adjust the action values of eligible relays.

\begin{table*}[t]
\caption{Main notation used throughout the paper.}\label{tab:notation}
\centering
\footnotesize
\setlength{\tabcolsep}{2.5pt}
\renewcommand{\arraystretch}{1.13}
\begin{tabular*}{\textwidth}{@{\extracolsep{\fill}}p{0.12\textwidth}p{0.38\textwidth}p{0.12\textwidth}p{0.38\textwidth}@{}}
\toprule
Symbol & Description & Symbol & Description\\
\midrule
\multicolumn{4}{@{}l}{\textbf{Reinforcement learning model and updates}}\\
$\mathcal S,\mathcal A$ & Network state space; global action space & $\mathcal P,\mathcal R$ & Transition mechanism and reward function\\
$\mathbf s_t^i,a_t^i$ & Local state and action used for execution and replay & $\mathcal A_t^i,\mathbf m_t^i$ & Candidate action set and binary mask\\
$a_t^{i,\mathrm{sel}},\widehat\pi$ & Selected action and policy retained for testing & $B_{\mathrm{mb}},\tau$ & Minibatch size and target update coefficient\\
$a_{\mathrm{sink}}$ & Direct-to-sink action ($N+1$) & $\mathbf s_t^{i,+},\mathbf m_t^{i,+},z_t$ & End-of-step representation, mask and termination flag\\
$r_t^i,r_t,J(\pi)$ & Head reward, mean reward per step and objective & $Q,V,A$ & Action value, state value and advantage\\
$\pi,\theta,\theta^-$ & Policy; online and target parameters & $\alpha,\gamma,\varepsilon$ & Q-learning rate, discount and exploration\\
$\mathcal D,\mathcal B$ & Replay buffer and minibatch & $\alpha_{\mathrm{PER}},\beta_{\mathrm{PER}},\omega_k$ & Priority/importance-sampling exponents; sample weight\\
$t,e,k$ & Time step, episode and replay record indices & $T_{\mathrm{ep}},e_{\max}$ & Step limit per episode and last allowable episode index\\
$h,\ell_k$ & Maximum and actual aggregation lengths & $\rho_t,\zeta$ & Progress within an episode and across training episodes\\
\midrule
\multicolumn{4}{@{}l}{\textbf{Network and energy models}}\\
$\mathcal V,N$ & Sensor-node set and count & $\mathcal C_t,\mathbf c_i$ & Cluster-head set and head $i$\\
$\mathcal K_t^i,K_t,q_t^i$ & Node candidates and counts before and after distance screening & $N_t^{\mathrm{alive}},n_t$ & Numbers of survivors before and after communication\\
$\mathbf v_n,(x_n,y_n)$ & Node $n$ and its position & $S,(x_S,y_S)$ & Sink and its position\\
$L_x,L_y,D_{\mathrm{area}}$ & Region dimensions and scale & $\mathbf D,d_{ij},d_{iS}$ & Distance matrix; node and sink distances\\
$E_0,E_n(t)$ & Initial energy and residual energy before a step & $d_0,d_{\max},d_{\mathrm{nbr}}$ & Crossover, relay-screening, neighborhood distances\\
$E_{\mathrm{elec}},e_{\mathrm{DA}}$ & Per-bit circuit and aggregation energy & $\varepsilon_{\mathrm{fs}},\varepsilon_{\mathrm{mp}}$ & Free-space/multipath amplifier coefficients\\
$L_{\mathrm{pkt}},B_t,B_{\mathrm{ref}}$ & Packet length; bits sent per step and reference & $E_{\mathrm{Tx}},E_{\mathrm{Rx}},E_{\mathrm{DA}}$ & Transmit, receive and aggregation energy\\
$C_{iS},C_{ij},C_{i\rightarrow j\rightarrow S}$ & Direct, single-hop relay and two-hop costs & $g_{ij},\kappa_{ij}$ & Progress gain and path-length ratio\\
$H_j,\nu_j,q_{\mathrm h}$ & Hotspot score; cumulative next-hop and update counts & $F_{ij}^{\mathrm{nbr}},F_{ij}^{\mathrm{fwd}}$ & Neighboring and forward candidate fractions\\
$\mu_t^+,\sigma_t^+$ & Mean and standard deviation of survivor energy after a step & $\overline{\mathrm{CV}}_t,w_t^x$ & Clipped energy CV and reward weight\\
$\eta_t,\chi_{ij}$ & Adaptive energy threshold and screening cost ratio & $B_{\mathrm{cum}}$ & Cumulative transmitted data\\
$P_{ij},\overline P_t^i$ & Action penalty and clipped reward penalty & $p_{\mathrm{ch}},T_t(n),\mathcal G_t^{\mathrm{CH}}$ & Head fraction, LEACH threshold and eligible set\\
$\mathcal G_t^{\mathrm{pkt}}$ & Packets counted as sent in time step $t$ & $C_t^i,C_{\mathrm{ref}}$ & Local and reference energy estimates\\
\botrule
\end{tabular*}
\end{table*}

\section{System Model and Problem Formulation}\label{sec:system-model}

This section describes the network and energy models, formulates the inter-cluster routing objective, and defines the state, action, and reward used for learning. Table~\ref{tab:notation} summarizes the main notation.

\subsection{Network Model}\label{subsec:network-model}

\begin{figure}[htbp]
\centering
\includegraphics[width=0.8\linewidth]{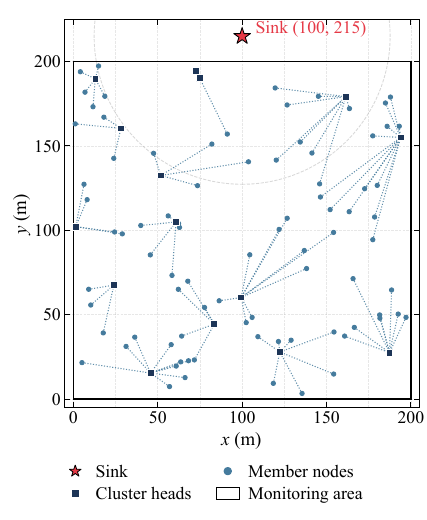}
\caption{An example of node locations and clustering in the WSN. The gray dashed circle is centered at the sink with radius $d_0$, the crossover distance in the radio energy model.}
\label{fig:network-overview}
\end{figure}

We consider a WSN with a static sink. Let $\mathcal V=\{\mathbf v_1,\mathbf v_2,\ldots,\mathbf v_N\}$ denote the sensor-node set. The nodes are independently deployed in a two-dimensional monitoring region, with positions drawn from uniform distributions:
\begin{equation}
\begin{aligned}
\mathbf v_n&=(x_n,y_n)^{\mathsf T},\\
x_n&\sim\operatorname{Unif}(0,L_x),\quad y_n\sim\operatorname{Unif}(0,L_y).
\end{aligned}
\label{eq:uniform-deployment}
\end{equation}
The sink $S$ is located outside the monitoring region. Node positions are drawn at the start of each training run and used throughout its training episodes and policy evaluations. Different seed settings yield different node configurations. In each time step, the environment forms clusters and assigns each member to its nearest cluster head. The reinforcement-learning methods use a common election rule. Fig.~\ref{fig:network-overview} shows the seed-42 configuration of 100 nodes in a $200\,\mathrm{m}\times200\,\mathrm{m}$ region, with the sink at $(100,215)\,\mathrm{m}$ and 15 heads in the illustrated clustering.

For nodes $\mathbf v_i$ and $\mathbf v_j$, the Euclidean distance is
\begin{equation}
d_{ij}=\sqrt{(x_i-x_j)^2+(y_i-y_j)^2},
\label{eq:node-distance}
\end{equation}
and the distance from node $\mathbf v_i$ to the sink is
\begin{equation}
d_{iS}=\sqrt{(x_i-x_S)^2+(y_i-y_S)^2}.
\label{eq:sink-distance}
\end{equation}
The pairwise distance matrix is constructed at initialization:
\begin{equation}
\mathbf D=
\begin{bmatrix}
0&d_{12}&\cdots&d_{1N}\\
d_{21}&0&\cdots&d_{2N}\\
\vdots&\vdots&\ddots&\vdots\\
d_{N1}&d_{N2}&\cdots&0
\end{bmatrix}.
\label{eq:distance-matrix}
\end{equation}
Member assignment compares distances to the current cluster heads. For inter-cluster routing, distances are used to rank and screen candidate relays and to describe their progress toward the sink and relative path lengths.

\subsection{Energy Consumption Model}\label{subsec:energy-model}

We use a first-order radio model~\cite{heinzelman2002application} to calculate transmission and reception energy. Transmitting a packet of $L_{\mathrm{pkt}}$ bits over distance $d$ consumes
\begin{equation}
E_{\mathrm{Tx}}(L_{\mathrm{pkt}},d)=
\begin{cases}
L_{\mathrm{pkt}}(E_{\mathrm{elec}}+\varepsilon_{\mathrm{fs}}d^2),&d\leq d_0,\\
L_{\mathrm{pkt}}(E_{\mathrm{elec}}+\varepsilon_{\mathrm{mp}}d^4),&d>d_0,
\end{cases}
\label{eq:transmit-energy}
\end{equation}
whereas reception consumes
\begin{equation}
E_{\mathrm{Rx}}(L_{\mathrm{pkt}})=L_{\mathrm{pkt}}E_{\mathrm{elec}}.
\label{eq:receive-energy}
\end{equation}
Here, $E_{\mathrm{elec}}$ is the circuit energy per bit, and $\varepsilon_{\mathrm{fs}}$ and $\varepsilon_{\mathrm{mp}}$ are the free-space and multipath amplifier coefficients, respectively. The crossover distance is $d_0=\sqrt{\varepsilon_{\mathrm{fs}}/\varepsilon_{\mathrm{mp}}}$. Data aggregation consumes
\begin{equation}
E_{\mathrm{DA}}(L_{\mathrm{pkt}})=L_{\mathrm{pkt}}e_{\mathrm{DA}},
\label{eq:data-aggregation}
\end{equation}
where $e_{\mathrm{DA}}$ is the aggregation energy per bit. The units of $E_{\mathrm{elec}}$ and $e_{\mathrm{DA}}$ are $\mathrm{J/bit}$, and those of $\varepsilon_{\mathrm{fs}}$ and $\varepsilon_{\mathrm{mp}}$ are $\mathrm{J}/(\mathrm{bit}\cdot\mathrm{m}^{2})$ and $\mathrm{J}/(\mathrm{bit}\cdot\mathrm{m}^{4})$, respectively. Cluster heads incur aggregation energy when receiving data from their members. During inter-cluster forwarding, the packets have already been aggregated, and residual energy is updated using transmission and reception costs.

For a cluster head $\mathbf c_i$, direct transmission to the sink costs
\begin{equation}
C_{iS}=E_{\mathrm{Tx}}(L_{\mathrm{pkt}},d_{iS}),
\label{eq:direct-cost}
\end{equation}
while transmission to a candidate relay $\mathbf c_j$ costs
\begin{equation}
C_{ij}=E_{\mathrm{Tx}}(L_{\mathrm{pkt}},d_{ij}).
\label{eq:relay-cost}
\end{equation}
If $\mathbf c_j$ receives the packet and forwards it directly to the sink, the two-hop communication cost is
\begin{equation}
\begin{aligned}
C_{i\rightarrow j\rightarrow S}={}&E_{\mathrm{Tx}}(L_{\mathrm{pkt}},d_{ij})
+E_{\mathrm{Rx}}(L_{\mathrm{pkt}})\\
&+E_{\mathrm{Tx}}(L_{\mathrm{pkt}},d_{jS}).
\end{aligned}
\label{eq:two-hop-cost}
\end{equation}
Repeated use of the same relay increases its communication burden and accelerates its energy depletion.

\subsection{Routing Problem Formulation}\label{subsec:routing-problem}

Given the clusters in a time step, the routing policy selects a next hop for each cluster head in sequence. Cluster heads retain their global node indices, so $\mathbf c_j=\mathbf v_j$ when node $j$ serves as a cluster head. The global action space is
\begin{equation}
	\mathcal A=\{1,\ldots,N,a_{\mathrm{sink}}\},\quad
	a_{\mathrm{sink}}=N+1,
\label{eq:global-action-space}
\end{equation}
where action $j$ selects node $\mathbf v_j$ and $a_{\mathrm{sink}}$ denotes direct transmission to the sink. After candidate-count and distance screening, the candidate action set for cluster head $i$ is
\begin{equation}
\mathcal A_t^i=\{j_1,\ldots,j_{q_t^i},a_{\mathrm{sink}}\}
\subseteq\mathcal A,
\label{eq:action-set}
\end{equation}
where $q_t^i$ is the number of retained node candidates. For $j\ne i$, action $j$ forwards the packet to cluster head $\mathbf c_j$. The self-node action $i$ is retained to represent direct transmission. When it is selected, both execution and the replay record use $a_{\mathrm{sink}}$, while retaining the penalty recorded during action selection. The network maintains separate outputs for action $i$ and $a_{\mathrm{sink}}$. The latter is always available. In the following, $a_t^i$ denotes the mapped index used for execution and replay.

The policy $\pi$ seeks to maximize the expected discounted step-level return:
\begin{equation}
J(\pi)=\mathbb E_\pi
\left[\sum_{t=0}^{T-1}\gamma^t r_t\right],
\label{eq:routing-objective}
\end{equation}
where $r_t$ is the mean reward across the cluster heads in time step $t$, $T$ is the number of completed time steps in the episode, and $T_{\mathrm{ep}}$ is the step limit per episode, with $T\leq T_{\mathrm{ep}}$. The reward combines node survival, transmitted data volume, communication energy efficiency, hotspot risk, and energy dispersion. Replay-based updates are described in Section~\ref{subsec:d3qn}. The routing decisions satisfy
\begin{subequations}\label{eq:routing-constraints}
\begin{align}
a_t^i&\in\mathcal A_t^i,\label{eq:constraint-action-feasibility}\\*
E_n(t+1)&=\max\{E_n(t)-\Delta E_n(t),0\},
\label{eq:constraint-energy-update}\\*
E_n(t+1)&\geq0,\quad\forall n,\label{eq:constraint-energy-nonnegative}\\*
a_t^i=j&\ \Rightarrow\ d_{ij}\leq d_{\max}.
\label{eq:constraint-relay-distance}
\end{align}
\end{subequations}
Here, $E_n(t)$ is the residual energy of node $n$ at the start of the time step, and $\Delta E_n(t)$ is its transmission, reception, and aggregation expenditure during that time step. The distance bound applies to node actions $j\in\{1,\ldots,N\}$; the direct-to-sink action is always available regardless of this relay-screening bound. Section~\ref{subsec:metrics} defines the transmitted-data metric, and Section~\ref{subsec:soft-action} describes the direction screening and soft penalties used in action selection.

To describe relay geometry, we define the progress gain and geometric path-length ratio as
\begin{equation}
\begin{aligned}
g_{ij}&=\frac{d_{iS}-d_{jS}}{\max(d_{iS},\varepsilon_d)},\quad \kappa_{ij}=\frac{d_{ij}+d_{jS}}{\max(d_{iS},\varepsilon_d)}.
\end{aligned}
\label{eq:routing-features}
\end{equation}
The positive constant $\varepsilon_d$ has units of distance and prevents division by zero. A positive $g_{ij}$ indicates that the relay is closer to the sink, while a smaller $\kappa_{ij}$ indicates a shorter relative path through that relay. Next-hop selection balances these geometric properties with communication cost, residual energy, and potential forwarding load.

\subsection{Markov Decision Process Formulation}\label{subsec:mdp}

We describe the inter-cluster next-hop selection problem using an MDP framework,
\begin{equation}
\mathcal M=(\mathcal S,\mathcal A,\mathcal P,\mathcal R,\gamma),
\label{eq:mdp}
\end{equation}
where $\mathcal S$ is the network state space, $\mathcal A$ is the global action space, $\mathcal P$ is the state-transition mechanism, $\mathcal R$ is the reward function, and $\gamma\in(0,1)$ is the discount factor. The network state includes the residual energies of all nodes and the communication progress. The agent uses a local representation of this state for each decision. The environment evolves in discrete time steps. In each time step, it first provides the cluster-head set and member assignments. The agent then selects a next hop for each cluster head. After all actions have been selected, the environment executes data transmission, updates energy, and calculates rewards.

\noindent\textbf{State.}
The local state representation $\mathbf s_t^i$ describes the relay conditions observed by cluster head $\mathbf c_i$. Node-feature blocks are concatenated in global node-index order, followed by a direct-to-sink block, giving $\mathbf s_t^i\in\mathbb R^{9(N+1)}$. For candidate relay $\mathbf c_j$, the feature block is
\begin{equation}
\begin{aligned}
\boldsymbol\phi_{ij}={}&[E_j,\mkern-2mu \bar E_j,\mkern-2mu \bar d_{jS},\mkern-2mu \bar d_{ij},\mkern-2mu H_j,\mkern-2mu g_{ij},\mkern-2mu \kappa_{ij},\mkern-2mu F_{ij}^{\mathrm{nbr}},\mkern-2mu F_{ij}^{\mathrm{fwd}}]^{\mathsf T}.
\end{aligned}
\label{eq:feature-block}
\end{equation}
The time-step index is omitted from the feature components for brevity. Here, $E_j$ is the relay's residual energy, $\bar E_j=E_j/E_0$ is its residual-energy ratio, and $E_0$ is the initial node energy. The sink distance $d_{jS}$ is normalized by $D_{\mathrm{area}}=\max(L_x,L_y)$, whereas the inter-node distance $d_{ij}$ is normalized by $d_0$, yielding $\bar d_{jS}$ and $\bar d_{ij}$, respectively.

The hotspot score represents how frequently a node appears in the retained next-hop records:
\begin{equation}
H_j=\min\left\{
\frac{c_{\mathrm{hot}}\nu_j}{\max(q_{\mathrm h},1)},1
\right\}.
\label{eq:hotspot-score}
\end{equation}
Here, $c_{\mathrm{hot}}$ scales the score. At each statistical update, $\nu_j$ increases by the number of retained cluster-head next-hop records pointing to node $j$, excluding self-node selections. The counter $q_{\mathrm h}$ records the number of updates. Both counts are reset at the start of each episode and updated before routing decisions in each time step.

The neighboring-candidate fraction $F_{ij}^{\mathrm{nbr}}$ and the forward-candidate fraction $F_{ij}^{\mathrm{fwd}}$ are calculated from the $q_t^i$ node candidates in $\mathcal A_t^i$. Let $\mathcal K_t^i$ be the set of these cluster heads, including $\mathbf c_i$, and define $M_i=\max(|\mathcal K_t^i|-1,1)$. For $j\ne i$,
\begin{equation}
\begin{aligned}
\mathcal N_{ij}&=\{\mathbf c_\ell\in\mathcal K_t^i\setminus\{\mathbf c_i,\mathbf c_j\}:d_{\ell j}\leq d_{\mathrm{nbr}}\},\\
\mathcal F_{ij}
&=\{\mathbf c_\ell\in\mathcal N_{ij}:d_{\ell S}\leq d_{jS}\},\\
F_{ij}^{\mathrm{nbr}}&=|\mathcal N_{ij}|/M_i,\qquad
F_{ij}^{\mathrm{fwd}}=|\mathcal F_{ij}|/M_i.
\end{aligned}
\label{eq:load-forward-features}
\end{equation}
The neighborhood distance $d_{\mathrm{nbr}}$ determines which candidates are counted. Thus, $F_{ij}^{\mathrm{nbr}}$ describes potential forwarding load through the relative number of nearby candidates, and $F_{ij}^{\mathrm{fwd}}$ describes the available candidates for onward forwarding toward the sink. Both use the number of candidates other than the current cluster head as their normalization basis.

Feature blocks for nodes outside the candidate list are zero. The direct-to-sink block is
\begin{equation}
[E_0,1,0,\bar d_{iS}^{\mathrm{sink}},0,1,1,0,1]^{\mathsf T},
\label{eq:sink-feature-block}
\end{equation}
where $\bar d_{iS}^{\mathrm{sink}}=d_{iS}/d_0$. The current cluster head retains its own feature block, with $F_{ii}^{\mathrm{nbr}}=0$. Its forward-candidate fraction $F_{ii}^{\mathrm{fwd}}$ is the number of other candidates within its neighborhood satisfying $d_{\ell S}\leq d_{iS}$, divided by $M_i$. Selecting its self-node action results in direct transmission to the sink. The feature bounds are given in Table~\ref{tab:params}.

\noindent\textbf{Action.}
The mapped action satisfies $a_t^i\in\mathcal A_t^i\subseteq\mathcal A$. Candidate availability varies with the surviving cluster heads and their distances, while the correspondence between network outputs and global actions remains fixed. A binary mask $\mathbf m_t^i\in\{0,1\}^{N+1}$ identifies the available actions: $m_{t,j}^i=1$ if and only if $j\in\mathcal A_t^i$, and $m_{t,a_{\mathrm{sink}}}^i=1$ at all times. Within the masked candidate set, the behavior policy applies direction and energy-efficiency screening and soft penalties, as described in Section~\ref{subsec:soft-action}. The selected next hops define the inter-cluster forwarding relations for the time step. A forwarding path follows these relations until a node has no next hop, points to itself, or points to a relay already recorded in the path; the terminal node then transmits directly to the sink.

\noindent\textbf{State transition.}
After transmission, node energy is updated and a successor state representation $\mathbf s_t^{i,+}$ and mask $\mathbf m_t^{i,+}$ are reconstructed using the surviving cluster heads. The current head's self-node entry is always retained in this reconstruction, even if its residual energy is zero. The superscript $+$ denotes this reconstruction at the end of the time step, which supplies the successor-state approximation used for replay learning. Decisions in the following time step use a newly formed cluster-head set. The successor mask is used in the Double-Q target. The termination flag is $z_t=1$ when all sensor nodes have failed and is zero otherwise. Reaching the step limit truncates the episode.

\noindent\textbf{Reward.}
The reward combines feedback on network operation with feedback on the current cluster head's action:
\begin{equation}
r_t^i=
\operatorname{clip}_{[r_{\min},r_{\max}]}
\left(r_t^{\mathrm{net}}+r_t^{i,\mathrm{loc}}\right),
\label{eq:reward}
\end{equation}
where $\operatorname{clip}_{[a,b]}(x)$ restricts $x$ to $[a,b]$, and $r_{\min}$ and $r_{\max}$ are the reward bounds. Network feedback is shared by all cluster heads within a time step, while local feedback depends on the selected action.

The network component is
\begin{equation}
\begin{aligned}
r_t^{\mathrm{net}}={}&
w_t^{\mathrm{sur}}\frac{n_t}{N}
+w_t^{\mathrm{dat}}\frac{B_t}{B_{\mathrm{ref}}}
-w_t^{\mathrm{bal}}\overline{\mathrm{CV}}_t,
\end{aligned}
\label{eq:reward-components}
\end{equation}
where $n_t$ is the number of surviving nodes at the end of the time step, $B_t$ is the total packet length counted during inter-cluster transmission, and $B_{\mathrm{ref}}$ is the data-volume normalization scale. The clipped coefficient of variation is
\begin{equation}
\overline{\mathrm{CV}}_t=
\min\left\{\frac{\sigma_t^+}{\max(\mu_t^+,\varepsilon_E)},1\right\},
\label{eq:clipped-energy-cv}
\end{equation}
where $\mu_t^+$ and $\sigma_t^+$ are the mean and standard deviation of the residual energy of surviving nodes at the end of the time step. Both are set to zero when all nodes have failed. The positive constant $\varepsilon_E$ has units of energy. The three terms reward node survival and transmitted data volume and penalize energy imbalance among surviving nodes.

The local component is
\begin{equation}
\begin{aligned}
r_t^{i,\mathrm{loc}}={}&
w^{\mathrm{eng}}\left(\frac{C_{\mathrm{ref}}}{C_t^i}-b_{\mathrm{eng}}\right)\\
&-w_t^{\mathrm{hot}}\widehat H_t^i
-w_t^{\mathrm{soft}}\overline P_t^i.
\end{aligned}
\label{eq:reward-local}
\end{equation}
The local energy estimate is
\begin{equation}
C_t^i=E_{\mathrm{Tx}}(L_{\mathrm{pkt}},d_t^i)
+E_{\mathrm{DA}}(L_{\mathrm{pkt}})>0,
\label{eq:local-energy-estimate}
\end{equation}
with $d_t^i=d_{ij}$ for a relay action and $d_t^i=d_{iS}$ for direct transmission. The reference energy is
\[
C_{\mathrm{ref}}=
E_{\mathrm{Tx}}(L_{\mathrm{pkt}},d_0)
+E_{\mathrm{DA}}(L_{\mathrm{pkt}}).
\]
It provides a common comparison scale at the radio-model crossover distance. These energy estimates are used to calculate the local energy-efficiency reward. The coefficient $w^{\mathrm{eng}}$ controls the energy-efficiency contribution, and $b_{\mathrm{eng}}$ is its baseline offset. This reward term increases as the local energy estimate decreases.

For a selected relay $\mathbf c_j$, $\widehat H_t^i=H_j$ and $\overline P_t^i=\min\{P_{ij},P_{\max}\}$, where $P_{ij}$ is the action-value penalty defined in Section~\ref{subsec:soft-action}. Both terms are zero for the independent direct-to-sink action. A selected self-node action is executed as direct transmission but retains the penalty recorded during action selection. Hotspot scores are taken from the statistics available before routing decisions in the time step, penalties are recorded during action selection, and rewards are calculated after transmission.

The energy-efficiency weight is fixed, while the other weights vary with communication progress within an episode. Define $\rho_t=\min\{t/T_{\mathrm{ep}},1\}$. For each time-varying weight,
\begin{equation}
w_t^x=(1-\rho_t)w_{\mathrm{start}}^x
+\rho_t w_{\mathrm{end}}^x,
\label{eq:reward-weight-schedule}
\end{equation}
where $x\in\{\mathrm{sur},\mathrm{dat},\mathrm{bal},\mathrm{hot},\mathrm{soft}\}$. The survival weight and the three penalty weights increase as communication progresses, while the transmitted-data weight decreases. The normalized reward components are dimensionless, and their weights need not sum to one. Table~\ref{tab:params} gives the weights and reward bounds.

Each cluster head contributes an experience record
\begin{equation}
(\mathbf s_t^i,a_t^i,r_t^i,\mathbf s_t^{i,+},z_t,\mathbf m_t^{i,+}).
\label{eq:experience-record}
\end{equation}
The action is stored using the mapped execution index, and the penalty assigned during action selection is incorporated into the reward. The step-level reward $r_t$ is the mean of $r_t^i$ across the cluster heads.

\section{HERO-D3QN for Inter-Cluster Routing}\label{sec:method}

\begin{figure}[!tp]
\centering
\includegraphics[width=0.9\columnwidth]{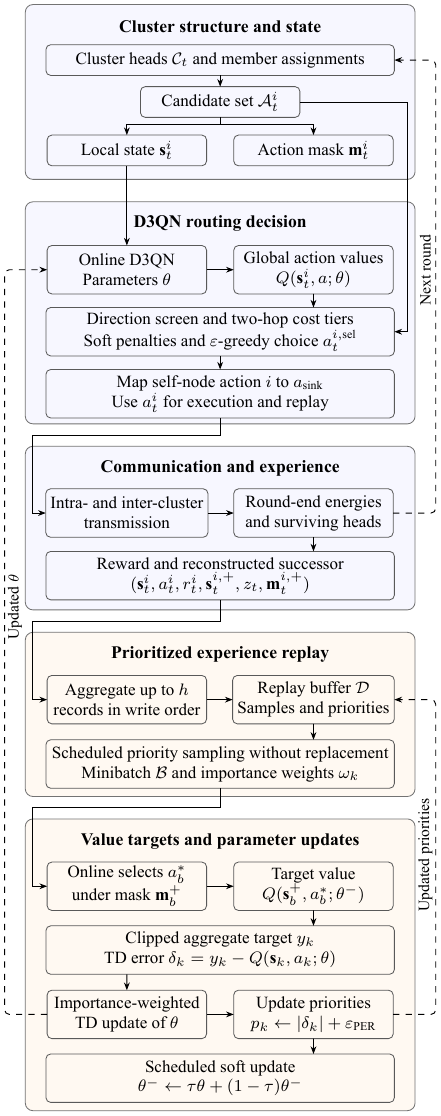}
\caption{HERO-D3QN training and routing framework. The current mask represents the candidate set; the reconstructed successor mask is used in the Double-Q target.}
\label{fig:diagram}
\end{figure}

HERO-D3QN learns a next-hop policy for the cluster structure formed in each time step. In Fig.~\ref{fig:diagram}, the candidate set $\mathcal A_t^i$ identifies the node entries used in the local state $\mathbf s_t^i$ and the available actions in the mask $\mathbf m_t^i$. The online D3QN estimates the action values. The selection rule screens relays by direction and forwarding cost and adjusts the values to account for low energy and excess cost. After selecting the next hops, the network transmits the data, updates node energies, and forms the experience records.

The learner aggregates the records in the order they enter a shared queue and samples the aggregates through prioritized experience replay (PER)~\cite{schaul2016prioritized}. The temporal-difference (TD) errors used in the online update also determine the new priorities of the sampled entries. The online network supplies the action values in subsequent routing decisions, and the target network is updated periodically.

\begin{figure}[t]
	\centering
	\includegraphics[width=\columnwidth]{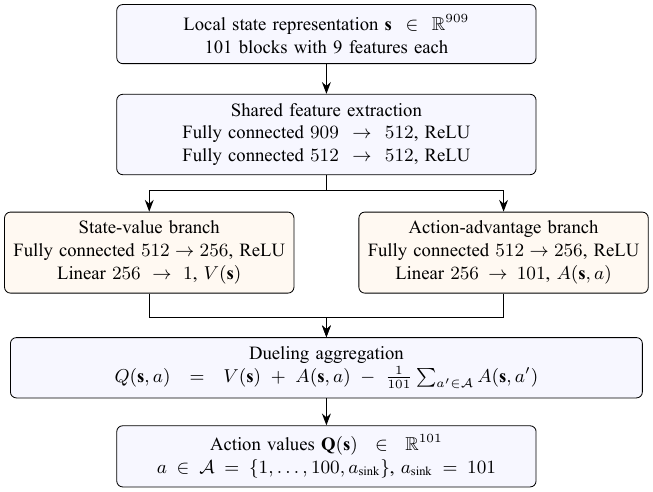}
	\caption{Shared D3QN architecture for $N=100$ with separate online and target parameters. Routing uses screened actions and soft penalties. Double-Q selection uses the successor mask.}
	\label{fig:d3qn-structure}
\end{figure}

\subsection{External Cluster Formation and Routing Interface}\label{subsec:cluster}

The environment provides the cluster-head set $\mathcal C_t$ and the member assignments before the routing decisions at each time step. Each surviving non-head node $n$ joins its nearest cluster head,
\begin{equation}
\mathbf c_t^*(n)=\operatorname*{arg\,min}_{\mathbf c\in\mathcal C_t}
d(\mathbf v_n,\mathbf c),
\label{eq:nearest-ch}
\end{equation}
where $d(\cdot,\cdot)$ is the Euclidean distance.

HERO-D3QN and the reinforcement-learning baselines use the same cluster-head election rule. Let $N_t^{\mathrm{alive}}$ be the number of surviving nodes at the start of time step $t$. When $N_t^{\mathrm{alive}}>0$, the environment selects $\max\{1,\lceil p_{\mathrm{ch}}N_t^{\mathrm{alive}}\rceil\}$ heads from the surviving nodes by sampling without replacement, with weights proportional to their residual energies. The LEACH baseline applies the election threshold in Eq.~\eqref{eq:leach-threshold} with a rotation cycle of $\max\{1,\lfloor 1/p_{\mathrm{ch}}\rfloor\}$ time steps and makes every surviving node eligible in each time step. It then randomly adds or removes heads to reach the count prescribed by the same formula for its surviving nodes.

For the reinforcement-learning routing methods, members within $d_{\mathrm{intra}}$ of their heads transmit first and send directly to them. Each more distant member forwards through the nearest surviving member of its cluster within $d_{\mathrm{intra}}$ of the head, or sends directly if no such member is available. In the LEACH baseline, members send directly to their heads. Table~\ref{tab:params} gives $p_{\mathrm{ch}}$ and $d_{\mathrm{intra}}$.

\subsection{D3QN Backbone}\label{subsec:d3qn}

For cluster head $\mathbf c_i$, the D3QN takes the local state representation $\mathbf s_t^i$ defined in Section~\ref{subsec:mdp}. Its $N+1$ outputs correspond to the global action space $\mathcal A$, with output $j$ for node action $j$ and the final output for $a_{\mathrm{sink}}$. In Fig.~\ref{fig:d3qn-structure}, the 909 input features for $N=100$ pass through two shared layers of width 512. The value and advantage branches each use 256 hidden units and produce one state value and 101 action advantages, respectively. The online and target networks have this architecture with separate parameters $\theta$ and $\theta^-$.

The dueling architecture combines the outputs of a state-value stream and an action-advantage stream~\cite{wang2016dueling}. For $a\in\mathcal A$, the online action value is
\begin{equation}
\begin{aligned}
Q(\mathbf s_t^i,a;\theta)
={}&V(\mathbf s_t^i;\theta)+A(\mathbf s_t^i,a;\theta)\\
&-\frac{1}{|\mathcal A|}
\sum_{a'\in\mathcal A} A(\mathbf s_t^i,a';\theta).
\end{aligned}
\label{eq:dueling-q}
\end{equation}
Here, $\theta$ includes the parameters of the shared layers and both streams. The mean advantage is computed over all $N+1$ outputs. Candidate masking and the selection rule in Section~\ref{subsec:soft-action} are applied to the resulting action values.

After each time step, the records from Section~\ref{subsec:mdp} enter a shared queue in cluster-head order. The replay index $k$ identifies a record in this sequence, giving the tuple $(\mathbf s_k,a_k,r_k,\mathbf s_k^+,z_k,\mathbf m_k^+)$. When the queue contains $h$ records, it produces a replay sample beginning with its oldest record and then removes that record. When a terminal record arrives, the remaining suffixes are stored as replay samples and the queue is cleared. At the step limit, records still in the queue carry over to the next episode. An aggregate can thus contain records from different heads and, across a truncation, from successive episodes.

For a sample beginning at record $k$, let $\ell_k\leq h$ be its actual aggregation length and $b=k+\ell_k-1$ its last record index. The successor state representation and mask are taken from record $b$. The successor action set is
\begin{equation}
\mathcal A_b^+
=\{j\mid 1\leq j\leq N,\ m_{b,j}^+=1\}
\cup\{a_{\mathrm{sink}}\}.
\label{eq:masked-action-set-target}
\end{equation}

The discounted reward for the aggregate is
\begin{equation}
\widehat R_k^{(\ell_k)}
=\sum_{q=0}^{\ell_k-1}\gamma^q r_{k+q}.
\label{eq:n-step-return}
\end{equation}
Here, each $r_{k+q}$ is the reward of the corresponding cluster head in the record sequence. This record-level approximation uses the reconstructed successor representation for learning. The discount in Eq.~\eqref{eq:n-step-return} counts records, while the routing objective in Section~\ref{subsec:routing-problem} discounts the mean reward $r_t$ by time step.

The Double-Q update uses the online network to select the successor action and the target network to evaluate it~\cite{hasselt2016double}. The online network selects
\begin{equation}
a_b^*=\operatorname*{arg\,max}_{a'\in\mathcal A_b^+}
Q(\mathbf s_b^+,a';\theta).
\label{eq:double-q-action}
\end{equation}
The target network evaluates the output indexed by $a_b^*$, giving the unclipped target
\begin{equation}
\widetilde y_k=\widehat R_k^{(\ell_k)}
+(1-z_b)\gamma^{\ell_k}Q(\mathbf s_b^+,a_b^*;\theta^-).
\label{eq:n-step-target}
\end{equation}
The sink action keeps $\mathcal A_b^+$ nonempty. The factor $1-z_b$ sets the bootstrap term to zero at network failure, and the exponent $\ell_k$ accounts for the actual number of records in the aggregate. The update uses $y_k=\operatorname{clip}_{[y_{\min},y_{\max}]}(\widetilde y_k)$, with the bounds given in Table~\ref{tab:params}.

\subsection{Energy-Aware Soft-Constrained Multihop Action Selection}\label{subsec:soft-action}

A relay can have a high estimated action value even when its energy is nearly depleted. HERO-D3QN screens node actions by direction and relative forwarding cost, then uses soft penalties to account for the energy and cost of the retained options. A relay below the adaptive energy threshold can still be selected when it meets the other criteria.

\noindent\textbf{Candidate relays and direction.}
For cluster head $\mathbf c_i$, the environment ranks the heads in $\mathcal C_t$ by their distance to $\mathbf c_i$ and considers the nearest $K_t$, where
\begin{equation}
K_t=\min\left\{|\mathcal C_t|,
\max\left(K_{\min},\lfloor \xi N\rfloor\right)\right\}.
\label{eq:candidate-cap}
\end{equation}
The fraction $\xi$ and the minimum $K_{\min}$ determine how many heads are considered before distance screening. Of these $K_t$ heads, those within $d_{\max}$ form the node candidate set $\mathcal K_t^i$, with $\mathbf c_i$ retained. Its size $q_t^i=|\mathcal K_t^i|$ is the number of node actions in Eq.~\eqref{eq:action-set}. Adding $a_{\mathrm{sink}}$ gives the full candidate action set $\mathcal A_t^i$.

Let $\mathbf p_{iS}=(x_i-x_S,y_i-y_S)^{\mathsf T}$ and $\mathbf p_{jS}=(x_j-x_S,y_j-y_S)^{\mathsf T}$ be the displacement vectors from the sink to $\mathbf c_i$ and $\mathbf c_j$. A node action passes direction screening if
\begin{equation}
\mathbf p_{iS}^{\mathsf T}\mathbf p_{jS}
\geq c_{\mathrm{dir}}\,
\lVert\mathbf p_{iS}\rVert_2\lVert\mathbf p_{jS}\rVert_2.
\label{eq:direction-constraint}
\end{equation}
The threshold $c_{\mathrm{dir}}\in[-1,1]$ sets the maximum angle between the two sink-centered vectors. The self-node action passes this screen.

\noindent\textbf{Adaptive energy threshold.}
Let $\mu_t$ and $\sigma_t$ be the mean and population standard deviation of the residual energies of the nodes alive before the routing decisions in time step $t$. Their counterparts after transmission, $\mu_t^+$ and $\sigma_t^+$, are used in the reward in Section~\ref{subsec:mdp}. The normalized threshold is
\begin{equation}
\eta_t=\max\left\{\eta_{\min},\frac{\mu_t-\lambda\sigma_t}{E_0}\right\}.
\label{eq:energy-threshold}
\end{equation}
Here, $\eta_{\min}$ is the threshold floor and $\lambda$ controls the influence of energy dispersion. At a given mean, greater dispersion lowers the threshold until the floor is reached. For node $j$, the normalized energy shortfall is
\begin{equation}
\Delta_{ij}^{E}
=\max\left\{\frac{\eta_t-\bar E_j}{\max(\eta_t,\varepsilon_\eta)},0\right\},
\label{eq:energy-gap}
\end{equation}
where $\bar E_j=E_j/E_0$ and $\varepsilon_\eta$ is a positive dimensionless constant.

Both soft penalties use a capped function of the normalized shortfall or excess $\Delta\geq0$ and the progress variable $\zeta\in[0,1]$,
\begin{equation}
\begin{aligned}
&\psi(\Delta,\zeta;\boldsymbol{\vartheta})\\
&\quad=\min\bigl\{\vartheta_5,
(\vartheta_1+\vartheta_2\Delta)(\vartheta_3+\vartheta_4\zeta)\bigr\}.
\end{aligned}
\label{eq:penalty-map}
\end{equation}
The five nonnegative entries of $\boldsymbol{\vartheta}$ set the base penalty, sensitivity to $\Delta$, initial multiplier, increase in that multiplier, and penalty cap. During training, $\zeta=\zeta_e=\min\{e/e_{\max},1\}$, where $e$ is the episode index and $e_{\max}$ is its upper bound. During evaluation, $\zeta=1$. This schedule varies the penalties across episodes, while $\rho_t$ in Section~\ref{subsec:mdp} varies the reward weights within each episode.

The low-energy penalty is
\begin{equation}
P_{ij}^{\mathrm{ene}}
=
\begin{cases}
\psi(\Delta_{ij}^{E},\zeta;\boldsymbol{\vartheta}^{\mathrm{ene}}),
& j\ne i,\ \bar E_j<\eta_t,\\
0,&\text{otherwise}.
\end{cases}
\label{eq:energy-soft-penalty}
\end{equation}
Thus, an energy shortfall lowers the adjusted action value; the relay remains eligible if it meets the other selection criteria. Table~\ref{tab:params} gives the penalty coefficients.

\noindent\textbf{Two-hop energy screening.}
The selection rule compares the energy cost of sending directly to the sink with the cost of forwarding through a candidate relay. It adds one aggregation term to each communication cost from Section~\ref{subsec:energy-model},
\begin{equation}
\begin{aligned}
\widehat C_{iS}&=C_{iS}+E_{\mathrm{DA}}(L_{\mathrm{pkt}}),\\
\widehat C_{i\rightarrow j\rightarrow S}
&=C_{i\rightarrow j\rightarrow S}+E_{\mathrm{DA}}(L_{\mathrm{pkt}}).
\end{aligned}
\label{eq:screening-cost}
\end{equation}
These estimates are used for screening. Inter-cluster energy expenditure is calculated from the transmission and reception costs in Section~\ref{subsec:energy-model}. The ratio used to screen node action $j$ is
\begin{equation}
\chi_{ij}=\frac{\widehat C_{i\rightarrow j\rightarrow S}}{\widehat C_{iS}}.
\label{eq:efficiency-ratio}
\end{equation}
For the self-node action, the same ratio is evaluated with $d_{ii}=0$.

When more than one node action passes the direction screen, the selector assigns candidates to two tiers.
\begin{itemize}
\item The priority tier contains candidates with $\chi_{ij}\leq1$ and $d_{jS}\leq d_{iS}$.
\item The relaxed tier contains the remaining candidates with $\chi_{ij}\leq\chi_{\max}$, where $\chi_{\max}>1$.
\end{itemize}

For a candidate in the relaxed tier, the normalized excess over the direct transmission cost is
\begin{equation}
\Delta_{ij}^{\chi}=\max\left\{\frac{\chi_{ij}-1}{\chi_{\max}-1},0\right\}.
\label{eq:efficiency-gap}
\end{equation}
The resulting energy-efficiency penalty is
\begin{equation}
P_{ij}^{\mathrm{eff}}
=
\begin{cases}
\psi(\Delta_{ij}^{\chi},\zeta;\boldsymbol{\vartheta}^{\mathrm{eff}}),
&\text{relaxed tier},\\
0,&\text{otherwise}.
\end{cases}
\label{eq:efficiency-soft-penalty}
\end{equation}
The selector uses the priority tier whenever it contains a candidate. If it is empty, the selector uses the relaxed tier when it contains at least two candidates. In the remaining cases, it uses the direction-screened set and retains any penalties already assigned. This fallback can retain a node action with $\chi_{ij}>\chi_{\max}$. The direct-to-sink action is added to the resulting set.

\noindent\textbf{Adjusted action values and execution.}
The two penalties are combined as
\begin{equation}
P_{ij}=P_{ij}^{\mathrm{ene}}+P_{ij}^{\mathrm{eff}},\qquad P_{iS}=0.
\label{eq:total-soft-penalty}
\end{equation}
The selector evaluates each retained action using
\begin{equation}
\begin{aligned}
\widetilde Q(\mathbf s_t^i,j;\theta)
&=Q(\mathbf s_t^i,j;\theta)-P_{ij},\\
\widetilde Q(\mathbf s_t^i,a_{\mathrm{sink}};\theta)
&=Q(\mathbf s_t^i,a_{\mathrm{sink}};\theta).
\end{aligned}
\label{eq:corrected-q}
\end{equation}

With probability $1-\varepsilon$, it selects an action with the largest adjusted value. With probability $\varepsilon$, it draws uniformly from the retained set. Let $a_t^{i,\mathrm{sel}}$ denote the action selected at this stage. If $a_t^{i,\mathrm{sel}}=i$, the executed and stored action is $a_t^i=a_{\mathrm{sink}}$. Otherwise, $a_t^i=a_t^{i,\mathrm{sel}}$. The penalty of the selected action enters the local reward in Section~\ref{subsec:mdp}.

\begin{algorithm*}[!t]
	\caption{HERO-D3QN training and routing}
	\label{alg:d3qn-routing}
	\small
	\begin{algorithmic}[1]
		\Require $\mathcal V$, $S$, and parameters in Section~\ref{subsec:settings}
		\Ensure Routing policy $\widehat\pi$
		\State Initialize $\theta$, $\varepsilon$, $\mathcal D$, and the shared queue. $\theta^-\gets\theta$, $n_{\mathrm{tr}}\gets0$.
		\For{$e=0,\ldots,e_{\max}$}
		\State Reset energies, routes, and hotspot counts. $\zeta\gets\min\{e/e_{\max},1\}$.
		\For{$t=0,\ldots,T_{\mathrm{ep}}-1$ \textbf{while} $\max_n E_n(t)>0$}
		\State Form $\mathcal C_t$ and clusters. Update $H_j$ and $\eta_t$.
		\For{$\mathbf c_i\in\mathcal C_t$}
		\State Build $(\mathcal A_t^i,\mathbf m_t^i,\mathbf s_t^i)$. Select $a_t^{i,\mathrm{sel}}$ using the rule in Section~\ref{subsec:soft-action}.
		\State Retain the selected action's penalty. $a_t^i\gets a_{\mathrm{sink}}$ \textbf{if} $a_t^{i,\mathrm{sel}}=i$, \textbf{else} $a_t^i\gets a_t^{i,\mathrm{sel}}$.
		\EndFor
		\State Execute the routes and obtain $(r_t^i,\mathbf s_t^{i,+},\mathbf m_t^{i,+},z_t)$.
		\State Append records in head order. Store aggregates by Eq.~\eqref{eq:n-step-return}, flushing suffixes at $z_t=1$.
		\If{$|\mathcal D|\geq2000$ \textbf{and} $n_{\mathrm{tr}}\bmod4=0$}
		\State Draw $\mathcal B$ by PER. Compute $y_k$, $\delta_k$, and $\omega_k$.
		\State Update $\theta$ using Eq.~\eqref{eq:loss}. $p_k\gets|\delta_k|+\varepsilon_{\mathrm{PER}}$, $k\in\mathcal B$.
		\If{$n_{\mathrm{tr}}\bmod200=0$}
		\State $\theta^-\gets\tau\theta+(1-\tau)\theta^-$.
		\EndIf
		\EndIf
		\State $n_{\mathrm{tr}}\gets n_{\mathrm{tr}}+1$.
		\EndFor
		\State Retain the queue at truncation. $\varepsilon\gets\max\{0.02,0.995\varepsilon\}$.
		\If{$e\bmod30=0$}
		\State Evaluate with $\varepsilon=0$ and $\zeta=1$.
		\If{$e>0$}
		\State Select $\widehat\pi$ using the criteria in Section~\ref{subsec:settings}. \textbf{if} the early stopping rule holds \textbf{then break}.
		\EndIf
		\EndIf
		\EndFor
		\State \Return $\widehat\pi$
	\end{algorithmic}
\end{algorithm*}

\subsection{Prioritized Replay and Parameter Updates}\label{subsec:training}

The replay buffer $\mathcal D$ stores each aggregate with the state and mapped action from its first record, the discounted reward in Eq.~\eqref{eq:n-step-return}, and the successor representation, mask, and termination flag from its last record. It also stores the actual aggregation length $\ell_k$. Let $\mathcal I_{\mathcal D}$ be the indices of the samples currently in the buffer and $M_{\mathcal D}=|\mathcal I_{\mathcal D}|$ its occupancy. For sample $k$ with priority $p_k$, the normalized sampling weight is
\begin{equation}
q_k=\frac{p_k^{\alpha_{\mathrm{PER}}}}{
\displaystyle\sum_{m\in\mathcal I_{\mathcal D}}p_m^{\alpha_{\mathrm{PER}}}},
\label{eq:per-prob}
\end{equation}
where $\alpha_{\mathrm{PER}}$ controls how strongly priority influences sampling. A new sample is assigned the larger of 1 and the highest priority currently in the buffer, with an initial priority of 1 when the buffer is empty.

A minibatch $\mathcal B$ is drawn without replacement using these weights. The importance-sampling weight for a sampled entry is computed from $q_k$ and normalized by the largest weight in the minibatch,
\begin{equation}
\omega_k=\frac{(M_{\mathcal D}q_k)^{-\beta_{\mathrm{PER}}}}{
\displaystyle\max_{m\in\mathcal B}(M_{\mathcal D}q_m)^{-\beta_{\mathrm{PER}}}},
\label{eq:is-weight}
\end{equation}
where $\beta_{\mathrm{PER}}$ is the importance-sampling exponent. For the clipped target $y_k$ in Section~\ref{subsec:d3qn}, the TD error is $\delta_k=y_k-Q(\mathbf s_k,a_k;\theta)$. The online update uses the weighted squared TD loss
\begin{equation}
\mathcal L(\theta)=\frac{1}{B_{\mathrm{mb}}}
\sum_{k\in\mathcal B}\omega_k\left[y_k-Q(\mathbf s_k,a_k;\theta)\right]^2,
\label{eq:loss}
\end{equation}
where $B_{\mathrm{mb}}=|\mathcal B|$. The targets are held fixed during the update. After the Adam step, each sampled entry receives the priority $p_k\leftarrow|\delta_k|+\varepsilon_{\mathrm{PER}}$, using the TD error computed for that update and a positive priority offset $\varepsilon_{\mathrm{PER}}$.

Learning begins once the replay buffer reaches its warm-up size. The online update intervals and the schedule for increasing $\beta_{\mathrm{PER}}$ are measured in cumulative training time steps. When a target update is due, it follows the online update and uses
\begin{equation}
\theta^-\leftarrow\tau\theta+(1-\tau)\theta^-,
\label{eq:soft-update}
\end{equation}
where $\tau$ is the soft update coefficient. Table~\ref{tab:params} gives the update intervals, gradient norm bound, and exploration parameters.

Each training episode starts by restoring node energies and resetting the routing state and hotspot statistics. It ends when all nodes have failed or the step limit is reached. Periodic evaluations measure network lifetime and cumulative transmitted data. The evaluation after episode $e=0$ provides a reference. Later evaluations inform policy selection and early stopping as described in Section~\ref{subsec:settings}. The retained parameters define the routing policy $\widehat\pi$ used in the final performance evaluation described in Section~\ref{sec:experiments}. Algorithm~\ref{alg:d3qn-routing} summarizes training and routing, with $n_{\mathrm{tr}}$ counting time steps across training episodes.

\begin{table*}[!t]
	\caption{Main simulation and learning parameters.}\label{tab:params}
	\centering
	\footnotesize
	\setlength{\tabcolsep}{3pt}
	\renewcommand{\arraystretch}{1.12}
	\begin{tabular*}{\textwidth}{@{\extracolsep{\fill}}p{0.25\textwidth}p{0.22\textwidth}p{0.25\textwidth}p{0.22\textwidth}@{}}
		\toprule
		Parameter & Value & Parameter & Value\\
		\midrule

		\textbf{Network and radio energy model}&&\textbf{Reward and soft penalties}&\\
		Monitoring region & $200\,\mathrm{m}\times200\,\mathrm{m}$ & Survival weight $w_t^{\mathrm{sur}}$ & $(1.00,1.35)$\\
		Number of nodes $N$ & 100 & Transmitted-data weight $w_t^{\mathrm{dat}}$ & $(1.05,0.70)$\\
		Sink position $(x_S,y_S)$ & $(100,215)\,\mathrm{m}$ & Balance weight $w_t^{\mathrm{bal}}$ & $(0.28,0.40)$\\
		Initial energy per node $E_0$ & $2\,\mathrm{J}$ & Hotspot weight $w_t^{\mathrm{hot}}$ & $(0.26,0.40)$\\
		Cluster-head fraction $p_{\mathrm{ch}}$ & 0.15 & Soft-penalty weight $w_t^{\mathrm{soft}}$ & $(0.162,0.324)$\\
		Packet length $L_{\mathrm{pkt}}$ & $8192\,\mathrm{bit}$ & Efficiency parameters $(w^{\mathrm{eng}},b_{\mathrm{eng}})$ & $(0.5,0.5)$\\
		Circuit energy $E_{\mathrm{elec}}$ & $50\,\mathrm{nJ/bit}$ & Transmitted-data scale $B_{\mathrm{ref}}$ & $10^5\,\mathrm{bit}$\\
		Free-space coefficient $\varepsilon_{\mathrm{fs}}$ & $10\,\mathrm{pJ}/(\mathrm{bit}\cdot\mathrm{m}^2)$ & Reward penalty cap $P_{\max}$ & 1.2\\
		Multipath coefficient $\varepsilon_{\mathrm{mp}}$ & $0.0013\,\mathrm{pJ}/(\mathrm{bit}\cdot\mathrm{m}^4)$ & Reward bounds $[r_{\min},r_{\max}]$ & $[-2,4]$\\
		Aggregation energy $e_{\mathrm{DA}}$ & $5\,\mathrm{nJ/bit}$ & Low-energy coefficients $\boldsymbol{\vartheta}^{\mathrm{ene}}$ & $(0.28,0.80,0.85,0.85,1.10)^{\mathsf T}$\\
		&  & Efficiency coefficients $\boldsymbol{\vartheta}^{\mathrm{eff}}$ & $(0.06,0.18,0.90,0.70,0.28)^{\mathsf T}$\\
		\midrule

		\textbf{Relay selection and state features}&&\textbf{HERO-D3QN learning}&\\
		Intra-cluster threshold $d_{\mathrm{intra}}$ & $75\,\mathrm{m}$ & Adam learning rate & $3\times10^{-4}$\\
		Candidate-count parameters $(\xi,K_{\min})$ & $(0.5,10)$ & Discount factor $\gamma$ & 0.99\\
		Relay distance limit $d_{\max}$ & $4d_0$ & Replay buffer capacity & $10^5$ samples\\
		Neighborhood distance $d_{\mathrm{nbr}}$ & $2.5d_0$ & Minimum buffer size for learning & $2000$ samples\\
		Hotspot scale $c_{\mathrm{hot}}$ & 5 & Minibatch size $B_{\mathrm{mb}}$ & 128\\
		Direction threshold $c_{\mathrm{dir}}$ & $-0.6$ ($126.9^\circ$) & Priority exponent $\alpha_{\mathrm{PER}}$ & 0.60\\
		Distance features $(\bar d_{jS},\bar d_{ij})$ & $[0,2],\ [0,5]$ & Importance-sampling exponent $\beta_{\mathrm{PER}}$ & $0.40$ to $1.00$ over $2\times10^5$ time steps\\
		Progress and path ratio $(g_{ij},\kappa_{ij})$ & $[-1,1],\ [0,3]$ & Priority offset $\varepsilon_{\mathrm{PER}}$ & $10^{-6}$\\
		Candidate fractions $(F_{ij}^{\mathrm{nbr}},F_{ij}^{\mathrm{fwd}})$ & $[0,1]$ each & Online and target update intervals & $4$ and $200$ time steps\\
		Sink distance ratio $\bar d_{iS}^{\mathrm{sink}}$ & $[0,10]$ & Soft update coefficient $\tau$ & 0.005\\
		Distance constant $\varepsilon_d$ & $10^{-9}\,\mathrm{m}$ & Exploration rate $\varepsilon$ & $0.90$ initially, minimum $0.02$\\
		Energy constant $\varepsilon_E$ & $10^{-9}\,\mathrm{J}$ & Exploration decay & Factor of $0.995$ per episode\\
		Threshold constant $\varepsilon_\eta$ & $10^{-9}$ & Maximum aggregation length $h$ & 3 records\\
		Adaptive threshold $(\lambda,\eta_{\min})$ & $(1.25,0.02)$ & Maximum gradient norm & 5\\
		Relaxed-tier cost ratio $\chi_{\max}$ & 1.30 & TD target bounds $[y_{\min},y_{\max}]$ & $[-100,100]$\\
		\bottomrule
	\end{tabular*}
\end{table*}

\section{Experimental Results and Analysis}\label{sec:experiments}

\subsection{Simulation Setup}\label{subsec:settings}

The comparison and ablation experiments use the same node configuration, generated with random seed 42, and a test horizon of 2,500 time steps. Table~\ref{tab:params} gives the simulation and learning parameters, Fig.~\ref{fig:d3qn-structure} shows the HERO-D3QN architecture, and Section~\ref{subsec:cluster} describes cluster formation and intra-cluster forwarding.

The feature intervals in Table~\ref{tab:params} give the clipping bounds for the state in Section~\ref{subsec:mdp}. For the time-varying reward weights, each pair gives the start and end values in Eq.~\eqref{eq:reward-weight-schedule}. The penalty coefficients follow the order in Eq.~\eqref{eq:penalty-map}. The update intervals and the linear increase in $\beta_{\mathrm{PER}}$ are measured in cumulative training time steps.

Each training episode lasts at most $T_{\mathrm{ep}}=2500$ time steps. HERO-D3QN and its variants use episode indices from 0 to $e_{\max}=2000$, with policy evaluations after episodes $e=0,30,60,\ldots$. The evaluation after episode 0 provides a reference, and model selection begins at $e=30$. The DQN and D3QN baselines are evaluated every 30 completed episodes and train for at most 2,000 episodes. Evaluations for model selection last up to 20,000 time steps. Both model selection and final testing use $\varepsilon=0$ for the learned methods. In these evaluations, the HERO-D3QN action selector uses $\zeta=1$.

For HERO-D3QN and its variants, model selection combines the first-node-death (FND), half-node-death (HND), and last-node-death (LND) indices with cumulative transmitted data. The four metrics are normalized against reference values updated across evaluations, with greater weight given to lifetime. A higher score or a qualifying gain in an individual metric can favor a new policy, subject to a check for persistent declines in FND, LND, or transmitted data. Training can end early when $e\geq300$ and five consecutive evaluations yield no qualifying improvement. For DQN and D3QN, model selection prioritizes FND, with HND and then LND used to break ties. Training ends early after five consecutive evaluations without improvement.

\subsection{Baselines and Performance Metrics}\label{subsec:metrics}

In the LEACH baseline, members send directly to their cluster heads, which in turn send directly to the sink. DQN, D3QN, and HERO-D3QN use the energy-weighted head election and intra-cluster forwarding described in Section~\ref{subsec:cluster}. The baseline labeled DQN uses a single-stream network with Double DQN updates, whereas D3QN uses a dueling network. Both represent each candidate action with five state features and select next hops from unadjusted action values, with candidate relays within $2.5d_0$. The hotspot feature is zero in the baseline tests. Their learning rule uses masked Double-Q targets and a reward that combines node survival with local energy cost.

FND, HND, and LND are the first indices at which the observed number of surviving nodes is below 100, at most 50, and zero, respectively. Time-step indices start at zero. In each time step, we record the number of surviving nodes before communication for LEACH, DQN, and D3QN, and after communication for HERO-D3QN and its variants.

After intra-cluster forwarding, the cluster heads are considered in sequence. For each head with positive residual energy at the start of its turn, we count one aggregated packet before deducting the energy cost of its route. Each packet is counted once at its originating head, irrespective of the number of relay hops. Let $\mathcal G_t^{\mathrm{pkt}}$ denote the packets counted in time step $t$ and $L_g$ the length of packet $g$ in bits. The data volume per step and its cumulative value are
\begin{equation}
B_t=\sum_{g\in\mathcal G_t^{\mathrm{pkt}}}L_g,
\qquad
B_{\mathrm{cum}}=\sum_t B_t.
\label{eq:transmitted-data}
\end{equation}
The reward in Eq.~\eqref{eq:reward-components} uses the same $B_t$. We report the last observed value of $B_{\mathrm{cum}}$ in Mbit, where $1\,\mathrm{Mbit}=10^6\,\mathrm{bit}$.

Total remaining energy is summed over all $N$ nodes. Energy dispersion is measured by the population variance of their residual energies, with zero energy for failed nodes. Peak and mean variances use all 2,500 observations, including the zeros after complete network failure. The five-seed summaries use the population standard deviation. The coefficient of variation (CV) is the ratio of this standard deviation to the mean.

\subsection{Network Lifetime and Energy Balance}\label{subsec:comparison}

Table~\ref{tab:comparison} summarizes the lifetime, transmission, and energy outcomes for the same node configuration.

\begin{table}[!tbp]
\caption{Network lifetime, transmitted data, and peak residual energy variance for the four routing methods (seed 42).}\label{tab:comparison}
\centering
\footnotesize
\setlength{\tabcolsep}{2pt}
\begin{tabular*}{\columnwidth}{@{\extracolsep{\fill}}lccccc@{}}
\toprule
Method & FND & HND & LND & Data (Mbit)\tnote{a} & Peak var. ($\mathrm{J}^2$)\tnote{b}\\
\midrule
LEACH & 441 & 1407 & $>2499$\tnote{c} & 176.26 & 0.2929\\
DQN & 229 & 1335 & 1863 & 164.46 & 0.2411\\
D3QN & 629 & 1356 & 2118 & 175.07 & 0.1783\\
HERO-D3QN & 1269 & 1378 & 1494 & 170.52 & 0.0793\\
\bottomrule
\end{tabular*}
\begin{tablenotes}
\footnotesize
\item[a] Cumulative transmitted data through the last observation.
\item[b] Maximum residual energy variance.
\item[c] LEACH's LND was not observed through time step 2499, the final observation.
\end{tablenotes}
\end{table}

\begin{figure}[!tbp]
\centering
\includegraphics[width=0.8\linewidth]{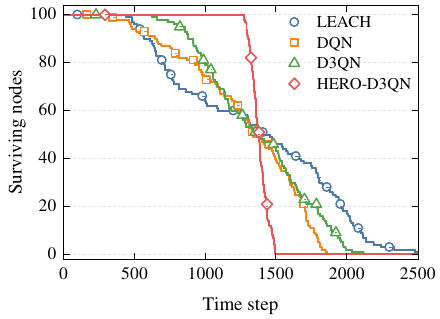}
\caption{Number of surviving nodes under different routing methods.}
\label{fig:comparison-lifetime}
\end{figure}

Fig.~\ref{fig:comparison-lifetime} compares node survival under the four routing methods. HERO-D3QN reached FND at time step 1269, approximately 2.88, 5.54, and 2.02 times the corresponding values for LEACH, DQN, and D3QN, respectively. Its HND of 1378 was later than those of DQN and D3QN but earlier than that of LEACH. HERO-D3QN reached LND at time step 1494, compared with 1863 for DQN and 2118 for D3QN. LEACH retained one surviving node at the last observation.
The interval from FND to LND was 225 time steps under HERO-D3QN, compared with 1634 for DQN and 1489 for D3QN, indicating a more concentrated failure process. These results show that the principal lifetime benefit of HERO-D3QN lies in maintaining the full node population for longer before failures begin. DQN and D3QN experienced earlier node failures followed by a more prolonged decline, which explains their later LNDs.

\begin{figure}[!tbp]
	\centering
	\includegraphics[width=0.8\linewidth]{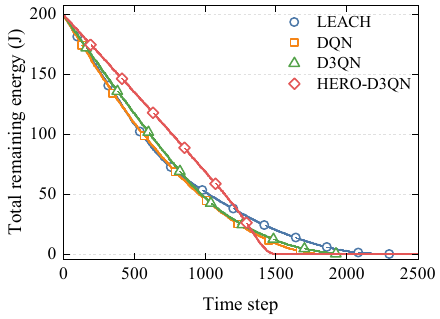}
	\caption{Total remaining energy under different routing methods.}
	\label{fig:comparison-energy}
\end{figure}

Fig.~\ref{fig:comparison-energy} shows the evolution of total remaining energy under the four routing methods. HERO-D3QN retained more energy than the baselines over most of the period before its first node failure. At time step 1000, it retained 68.9~J, compared with 51.9~J for LEACH, 45.9~J for DQN, and 47.2~J for D3QN. This larger reserve is consistent with slower early energy depletion and accompanies the delayed first node failure shown in Fig.~\ref{fig:comparison-lifetime}.
The remaining energy under HERO-D3QN reached zero at time step 1494, while the baselines still retained positive energy. Their curves subsequently declined more gradually as their surviving node populations decreased. The longer baseline tails thus reflect the persistence of energy in a reduced set of surviving nodes. Together, Figs.~\ref{fig:comparison-lifetime} and~\ref{fig:comparison-energy} distinguish energy preservation before the onset of node failures from residual-energy persistence after substantial node loss.

\begin{figure}[!tbp]
\centering
\includegraphics[width=0.8\linewidth]{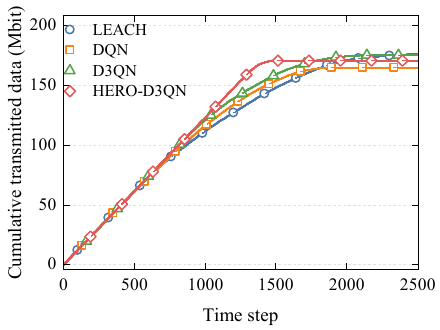}
\caption{Cumulative transmitted data under different routing methods.}
\label{fig:comparison-throughput}
\end{figure}

Fig.~\ref{fig:comparison-throughput} compares cumulative transmitted data under the four routing methods. The curves initially increase at similar rates, but HERO-D3QN subsequently accumulates more data before reaching its plateau. At time step 1269, its cumulative volume was 156.06~Mbit, compared with 132.18, 139.12, and 144.02~Mbit for LEACH, DQN, and D3QN, respectively. This separation is consistent with the delayed onset of node failures in Fig.~\ref{fig:comparison-lifetime}.
Under HERO-D3QN, cumulative transmitted data reached 170.52~Mbit at time step 1494 and remained at this level thereafter. D3QN and LEACH continued transmitting and surpassed this volume at time steps 1788 and 1986, reaching 175.07 and 176.26~Mbit, respectively, by the last observation. DQN ended at 164.46~Mbit, 6.06~Mbit below HERO-D3QN. The final ordering therefore reflects both the data accumulated before extensive node loss and the additional transmissions sustained by the baselines' surviving nodes.

\begin{figure}[!tbp]
\centering
\includegraphics[width=0.8\linewidth]{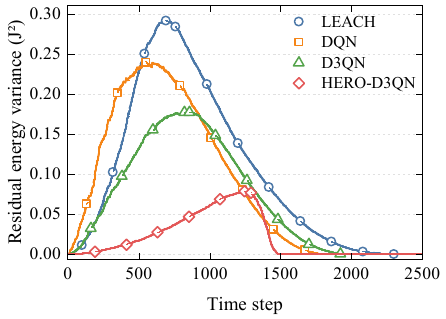}
\caption{Residual energy variance under different routing methods.}
\label{fig:comparison-variance}
\end{figure}

The residual energy variance in Fig.~\ref{fig:comparison-variance} rose more gradually under HERO-D3QN and peaked at $0.0793\,\mathrm{J}^2$ at time step 1245. This peak was approximately 72.9\%, 67.1\%, and 55.5\% below those of LEACH, DQN, and D3QN, respectively. The corresponding baseline maxima occurred at time steps 692, 551, and 857.
All 100 nodes remained alive at the HERO-D3QN peak, whereas node failures preceded the baseline maxima. The lower dispersion thus accompanied the longer period of full node survival in Fig.~\ref{fig:comparison-lifetime}. Variance then decreased with the depletion of total remaining energy shown in Fig.~\ref{fig:comparison-energy}, reaching zero at time step 1494 for HERO-D3QN when its last node failed. The baseline curves declined over longer survival tails.

\begin{figure*}[tp]
\centering
\subfloat[Lifetime endpoints]{
	\includegraphics[width=0.3\textwidth]{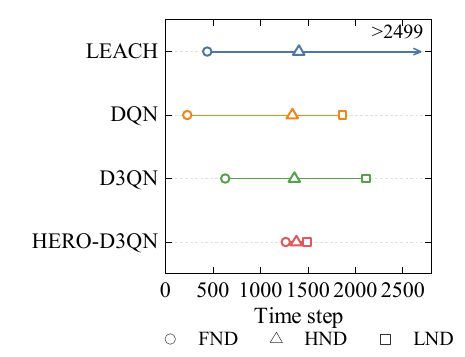}
	\label{fig:comparison-endpoints-a}
	}
\subfloat[Cumulative volume]{
	\includegraphics[width=0.3\textwidth]{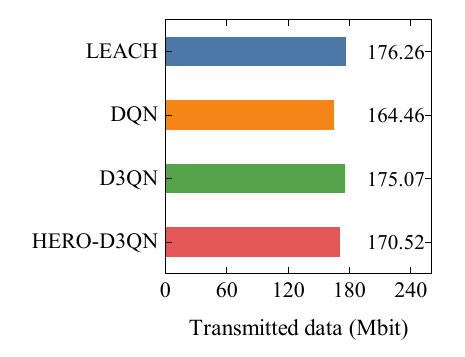}
	\label{fig:comparison-endpoints-b}
	}
\subfloat[Energy balance]{
	\includegraphics[width=0.3\textwidth]{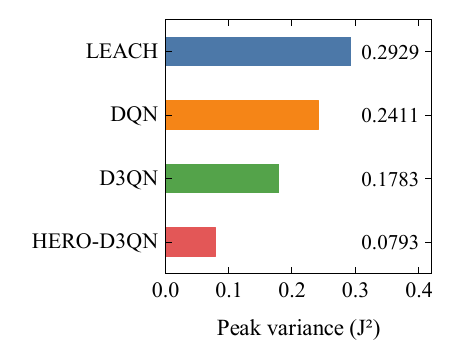}
	\label{fig:comparison-endpoints-c}
	}
\caption{Comparison of (a) lifetime endpoints, (b) cumulative transmitted data at the last observation, and (c) peak residual energy variance. Lines in (a) span FND to LND; the arrow denotes LEACH's unobserved LND ($>2499$).}
\label{fig:comparison-endpoints}
\end{figure*}

The HND indices of the three learned methods in Fig.~\ref{fig:comparison-endpoints-a} differed by only 43 time steps, despite their distinct patterns of node loss. Figs.~\ref{fig:comparison-endpoints-b} and~\ref{fig:comparison-endpoints-c} show that HERO-D3QN had the lowest peak residual energy variance and a final transmitted volume higher than DQN's but lower than those of LEACH and D3QN.

\begin{figure*}[tp]
\centering
\subfloat[Policy learning]{
	\includegraphics[width=0.35\textwidth]{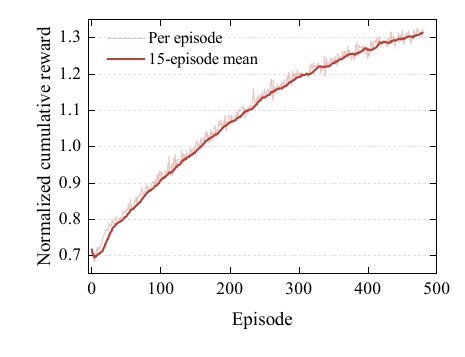}
	}
\subfloat[Value estimation]{
	\includegraphics[width=0.35\textwidth]{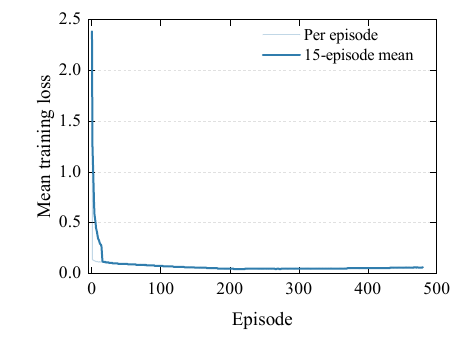}
	}
\caption{Training curves for HERO-D3QN: (a) normalized cumulative reward and (b) mean training loss. Thin lines show individual episodes, and thick lines show trailing means over up to 15 episodes.}
\label{fig:training-curves}
\end{figure*}

Fig.~\ref{fig:training-curves} shows the seed-42 training run over 481 episodes. For each episode, the normalized cumulative reward is obtained by summing the mean reward at each time step and dividing by $T_{\mathrm{ep}}$. The mean training loss is the weighted TD loss in Eq.~\eqref{eq:loss}, averaged over the updates in that episode. Because the normalization horizon is fixed, the normalized cumulative reward also reflects episode duration. The 15-episode mean reward increased from 0.710 at $e=14$ to 1.315 at $e=480$, with smaller gains late in training.
Mean training loss fell sharply from 2.385 at $e=0$ to 0.140 at $e=1$, then decreased more gradually. During the later episodes ($e=200,\ldots,480$), the loss exhibited small episode-to-episode fluctuations within approximately 0.04--0.07.

\subsection{Ablation Study}\label{subsec:ablation}

The variants follow the HERO-D3QN training settings for seed 42. In \emph{Base action chooser}, cluster heads select from the candidate action set $\mathcal A_t^i$ using unadjusted action values. This replaces direction screening, cost screening, and soft penalties. \emph{No hotspot penalty} removes the hotspot term from the reward and retains the hotspot feature in the state. \emph{Basic state} assigns the constants $0,1,0,0$ to $g_{ij}$, $\kappa_{ij}$, $F_{ij}^{\mathrm{nbr}}$, and $F_{ij}^{\mathrm{fwd}}$, respectively, in the node feature blocks. With \emph{Direct intra-cluster transmission}, members send directly to their heads, which use HERO-D3QN for inter-cluster routing. The full HERO-D3QN row in Table~\ref{tab:ablation} uses the result in Table~\ref{tab:comparison}.

\begin{table}[!tbp]
\caption{Network lifetime and transmitted data for HERO-D3QN and its variants (seed 42).}\label{tab:ablation}
\centering
\footnotesize
\setlength{\tabcolsep}{2pt}
\begin{tabular*}{\columnwidth}{@{\extracolsep{\fill}}lcccc@{}}
\toprule
Configuration & FND & HND & LND & Data (Mbit)\tnote{a}\\
\midrule
HERO-D3QN & 1269 & 1378 & 1494 & 170.52\\
Base action chooser & 275 & 1036 & 1830 & 133.88\\
No hotspot penalty & 717 & 913 & 1024 & 112.84\\
Basic state & 876 & 1106 & 1165 & 134.55\\
Direct intra-cluster & 974 & 1137 & 1229 & 139.39\\
\bottomrule
\end{tabular*}
\begin{tablenotes}
\footnotesize
\item[a] Cumulative transmitted data through the last observation.
\end{tablenotes}
\end{table}

\begin{figure}[tbp]
\centering
\includegraphics[width=0.8\linewidth]{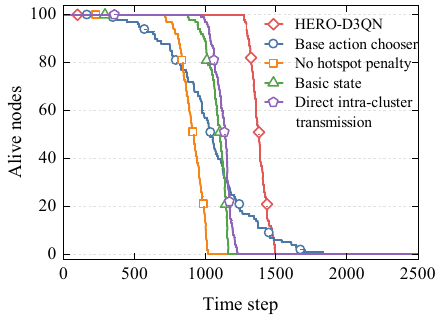}
\caption{Node survival for HERO-D3QN and its variants.}
\label{fig:ablation-lifetime}
\end{figure}

The node survival curves in Fig.~\ref{fig:ablation-lifetime} show that HERO-D3QN reached FND and HND at time steps 1269 and 1378, respectively. Across the four variants, FND ranged from 275 to 974 and HND from 913 to 1137. At time step 1269, HERO-D3QN had 99 surviving nodes, compared with 17 for the base action chooser and none for the other three variants.
The base action chooser reached LND at time step 1830, but only six nodes remained alive at time step 1494, when HERO-D3QN reached LND. Its later LND therefore reflected the persistence of a few nodes after extensive node loss.

\begin{figure}[!tbp]
\centering
\includegraphics[width=0.8\linewidth]{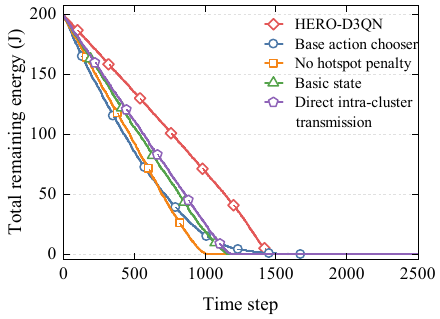}
\caption{Total remaining energy for HERO-D3QN and its variants.}
\label{fig:ablation-energy}
\end{figure}

Fig.~\ref{fig:ablation-energy} compares total remaining energy under HERO-D3QN and its variants. At time step 1000, HERO-D3QN retained 68.9~J, while the four variants retained 0.7--25.1~J. Direct intra-cluster transmission had the largest reserve among the variants, with 25.1~J and 97 surviving nodes, compared with 100 under HERO-D3QN. The higher energy reserve under HERO-D3QN thus accompanied the longer period of full node survival in Fig.~\ref{fig:ablation-lifetime}.
At time step 1494, the network under HERO-D3QN had exhausted its energy, while the network using the base action chooser retained 0.84~J in six surviving nodes and did not reach complete energy depletion until time step 1830.

\begin{figure}[!tbp]
\centering
\includegraphics[width=0.8\linewidth]{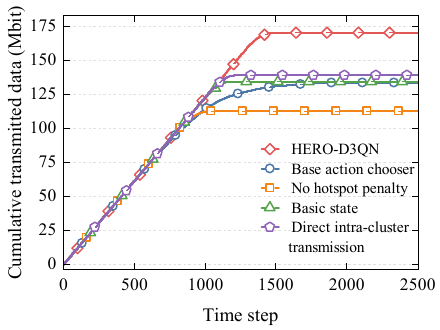}
\caption{Cumulative transmitted data for HERO-D3QN \mbox{and its variants}.}
\label{fig:ablation-throughput}
\end{figure}

Fig.~\ref{fig:ablation-throughput} compares cumulative transmitted data under HERO-D3QN and its variants. By time step 1269, when its first node failed, HERO-D3QN had accumulated 156.06~Mbit, already exceeding the final total of every variant. Its final volume reached 170.52~Mbit, compared with 112.84--139.39~Mbit for the four variants. The highest variant total was obtained with direct intra-cluster transmission.
The base action chooser continued transmitting until time step 1830, adding 2.75~Mbit after HERO-D3QN reached its plateau at time step 1494. The additional transmissions during the extended survival tail in Fig.~\ref{fig:ablation-lifetime} narrowed the gap, but the base action chooser's final volume of 133.88~Mbit remained 36.64~Mbit below that of HERO-D3QN.

\begin{figure}[!tbp]
\centering
\includegraphics[width=0.8\linewidth]{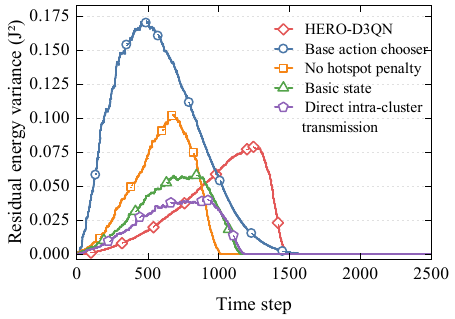}
\caption{Residual energy variance for HERO-D3QN \mbox{and its variants}.}
\label{fig:ablation-variance}
\end{figure}

Fig.~\ref{fig:ablation-variance} shows the evolution of residual energy variance under HERO-D3QN and its variants. Under HERO-D3QN, variance peaked at $0.0793\,\mathrm{J}^2$ at time step 1245, 53.8\% below the $0.1717\,\mathrm{J}^2$ peak of the base action chooser at time step 489. All 100 nodes remained alive at the HERO-D3QN peak, compared with 97 at the base action chooser's peak. The variant without the hotspot term in the reward peaked at $0.1031\,\mathrm{J}^2$ at time step 672, while all 100 nodes remained alive. Its greater energy dispersion relative to HERO-D3QN was therefore already evident before node failure.
The basic-state variant and the variant with direct intra-cluster transmission had lower peak variances of $0.0583$ and $0.0406\,\mathrm{J}^2$, respectively. These lower peaks accompanied earlier node failures and smaller transmitted volumes in Figs.~\ref{fig:ablation-lifetime} and~\ref{fig:ablation-throughput}. Their variance curves reached zero at time steps 1165 and 1229, compared with 1494 for HERO-D3QN and 1830 for the base action chooser. At these endpoints, all node energies were zero, as shown in Fig.~\ref{fig:ablation-energy}, so the late decrease in variance also reflected energy depletion.

\begin{figure*}[!tp]
\centering
\subfloat[Lifetime endpoints]{
	\includegraphics[width=0.3\textwidth]{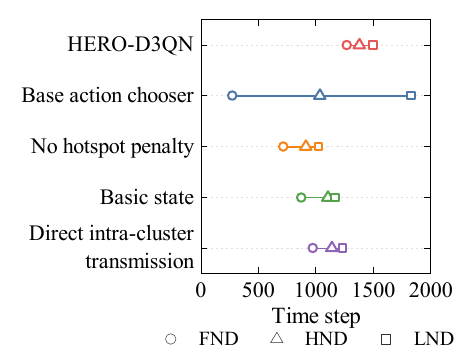}
	\label{fig:ablation-endpoints-a}
	}
\subfloat[Cumulative volume]{
	\includegraphics[width=0.3\textwidth]{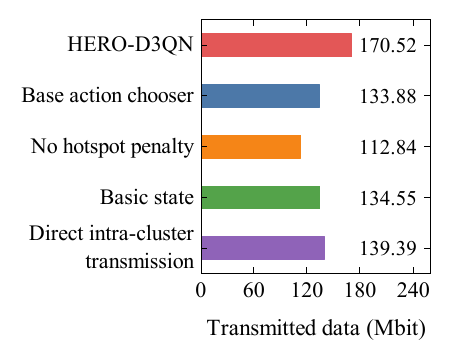}
	\label{fig:ablation-endpoints-b}
	}
\subfloat[Energy balance]{
	\includegraphics[width=0.3\textwidth]{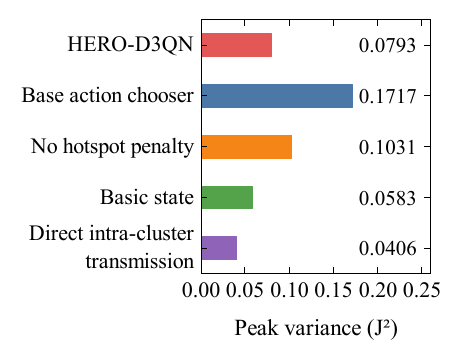}
	\label{fig:ablation-endpoints-c}
	}
\caption{Lifetime, transmitted data, and energy dispersion for HERO-D3QN and its variants. Lines in (a) span FND to LND, with HND marked between them. Panels (b) and (c) show cumulative transmitted data at the last observation and peak residual energy variance, respectively.}
\label{fig:ablation-endpoints}
\end{figure*}

Figs.~\ref{fig:ablation-endpoints-a} and~\ref{fig:ablation-endpoints-b} show that HERO-D3QN combined the latest FND and HND with the largest cumulative transmitted volume among the variants. The base action chooser had the latest LND, with 1555 time steps between FND and LND, but its earlier FND and HND were accompanied by a lower transmitted volume than HERO-D3QN. In Fig.~\ref{fig:ablation-endpoints-c}, the basic-state variant and the variant with direct intra-cluster transmission had lower peak energy variances than HERO-D3QN, despite their earlier node failures and smaller transmitted volumes.

\subsection{Routing Patterns and Seed Variability}\label{subsec:stability}

\begin{figure*}[!tp]
	\centering
	\subfloat[Training, $e=5$]{
		\includegraphics[width=0.25\textwidth]{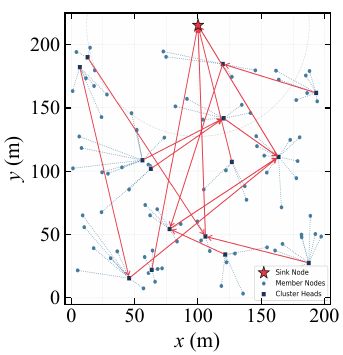}
		\label{fig:topology-snapshots-a}
		}
	\subfloat[Training, $e=90$]{
		\includegraphics[width=0.25\textwidth]{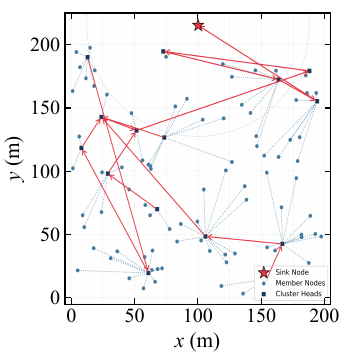}
		\label{fig:topology-snapshots-b}
		}
	\subfloat[Model selection]{
		\includegraphics[width=0.25\textwidth]{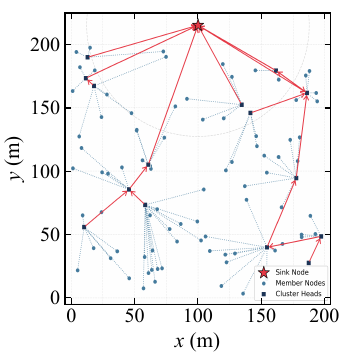}
		\label{fig:topology-snapshots-c}
		}
	\par\smallskip
	\includegraphics[width=\textwidth]{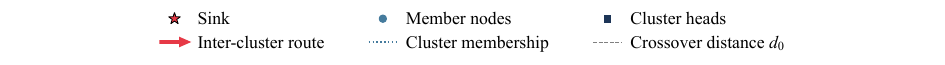}
	\caption{Inter-cluster routes for seed 42 at (a) training episode $e=5$ with $\varepsilon=0.88$, (b) training episode $e=90$ with $\varepsilon=0.57$, and (c) a model-selection evaluation. Arrows follow the inter-cluster routes, dotted lines show member assignments, and the dashed arc marks the crossover distance $d_0$ from the sink. Coordinates are in meters.}
	\label{fig:topology-snapshots}
\end{figure*}

Fig.~\ref{fig:topology-snapshots} shows three clustering configurations and their inter-cluster routes for seed 42. The first two snapshots show training episodes 5 and 90, with exploration rates of 0.88 and 0.57. The third shows routes formed during model selection. The cluster-head selections, member assignments, relay paths, and path crossings differ across the three snapshots. Because candidate relays are drawn from the current set of cluster heads, the different clusterings change the forwarding choices available to each head.

The five seed settings jointly vary node locations and the random processes used in training and evaluation. Each test uses the policy trained under its corresponding seed and spans 2,500 time steps. The seed-42 entry in Table~\ref{tab:seed} uses the HERO-D3QN result reported in Tables~\ref{tab:comparison} and~\ref{tab:ablation}.

\begin{table*}[!t]
\begin{threeparttable}
\caption{Network lifetime, transmitted data, and peak and time-averaged residual energy variance for HERO-D3QN across five seed settings.}\label{tab:seed}
\centering
\footnotesize
\begin{tabular*}{\textwidth}{@{\extracolsep{\fill}}lcccccc@{}}
\toprule
%Seed & FND & HND & LND & \shortstack{Transmitted\\data (Mbit)} & \shortstack{Peak variance\\($\mathrm{J}^2$)} & \shortstack{Mean variance\\($\mathrm{J}^2$)}\\

Seed & FND & HND & LND & Transmitted data (Mbit)\tnote{a} & Peak variance ($\mathrm{J}^2$)\tnote{b} & Mean variance ($\mathrm{J}^2$)\tnote{c}\\
\midrule
7 & 1268 & 1387 & 1879 & 176.54 & 0.0688 & 0.0193\\
21 & 1009 & 1438 & 2077 & 183.47 & 0.1049 & 0.0347\\
42 & 1269 & 1378 & 1494 & 170.52 & 0.0793 & 0.0205\\
84 & 1199 & 1319 & 1456 & 164.12 & 0.1016 & 0.0246\\
168 & 1258 & 1390 & 2209 & 179.22 & 0.0870 & 0.0233\\
Mean $\pm$ SD\tnote{d} & $1200.6\pm99.2$ & $1382.4\pm38.0$ & $1823.0\pm303.2$ & $174.77\pm6.78$ & $0.0883\pm0.0135$ & $0.0245\pm0.0055$\\
\bottomrule
\end{tabular*}
\begin{tablenotes}
\footnotesize
\item[a] Cumulative transmitted data through the last observation.
\item[b] Maximum residual energy variance over all 2,500 time steps.
\item[c] Residual energy variance averaged over all 2,500 time steps, including the zeros after complete network failure.
\item[d] The mean and population standard deviation are computed across the five seed settings.
\end{tablenotes}
\end{threeparttable}
\end{table*}

\begin{figure*}[!tp]
	\centering
	\subfloat[Lifetime endpoints]{
		\includegraphics[width=0.25\textwidth]{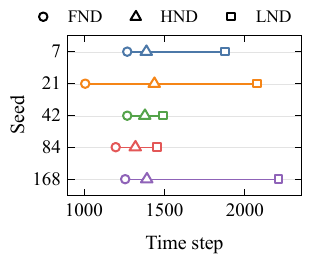}\label{fig:seed-metric-distribution-a}
	}
	\subfloat[Cumulative volume]{
		\includegraphics[width=0.25\textwidth]{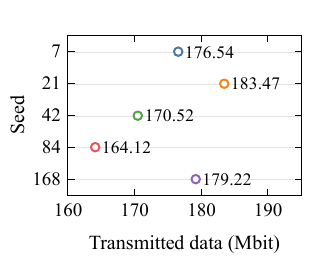}
		\label{fig:seed-metric-distribution-b}
	}
	\subfloat[Energy dispersion]{
		\includegraphics[width=0.25\textwidth]{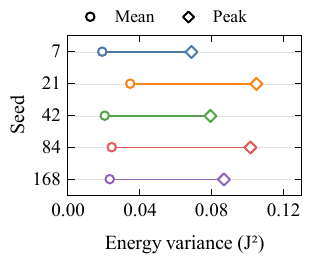}
		\label{fig:seed-metric-distribution-c}
	}
	\caption{HERO-D3QN outcomes across five seed settings: (a) node-failure endpoints, (b) cumulative transmitted data, and (c) time-averaged and peak residual energy variance. Rows and colors identify the same seeds across panels. Lines in (a) span FND to LND, with HND marked between them; lines in (c) connect the two variance measures for each seed.}
	\label{fig:seed-metric-distribution}
\end{figure*}

Fig.~\ref{fig:seed-metric-distribution-a} shows that HND varied less across the five seed settings than FND or LND. The CVs of FND, HND, and LND were 8.26\%, 2.75\%, and 16.63\%, respectively. Seed 21 reached FND earliest, at time step 1009, and HND latest, at time step 1438. Its interval from FND to LND was 1068 time steps, compared with 225 and 257 for seeds 42 and 84. The timing of the first failure alone therefore did not capture the later progression of node loss.
Fig.~\ref{fig:seed-metric-distribution-b} shows cumulative transmitted volumes from $164.12$ to $183.47\,\mathrm{Mbit}$, with a CV of 3.88\%. Seed 21 had the largest volume and the highest peak and time-averaged energy variances in Fig.~\ref{fig:seed-metric-distribution-c}, $0.1049$ and $0.0347\,\mathrm{J}^2$, respectively. The CVs of these two variance measures were 15.34\% and 22.29\%, respectively. Relative differences in energy dispersion were therefore larger than those in HND and transmitted volume. Seed 168 reached the latest LND, at time step 2209, with both variance indicators below their five-seed means.

\begin{figure}[!tbp]
	\centering
	\includegraphics[width=0.8\linewidth]{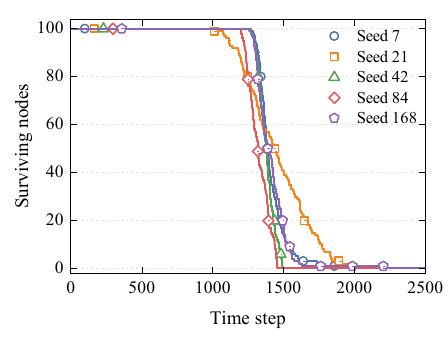}
	\caption{Node survival under HERO-D3QN across five seed settings.}
	\label{fig:seed-alive}
\end{figure}

Fig.~\ref{fig:seed-alive} shows that the relatively close HND values were reached through different patterns of node loss. The interval from FND to HND was 109 time steps under seed 42, compared with 429 under seed 21. Seed 42's longer initial period without failures was thus followed by a more concentrated decline, while seed 21 lost nodes more gradually during this phase.
At time step 1600, seed 21 retained 29 nodes and seed 168 only three, yet seed 168 reached the latest LND, at time step 2209. Its later LND reflected the persistence of a very small surviving population. The greater variation in LND than in HND shown in Fig.~\ref{fig:seed-metric-distribution-a} is consistent with these differences in the survival tails.

\begin{figure}[!tbp]
	\centering
	\includegraphics[width=0.8\linewidth]{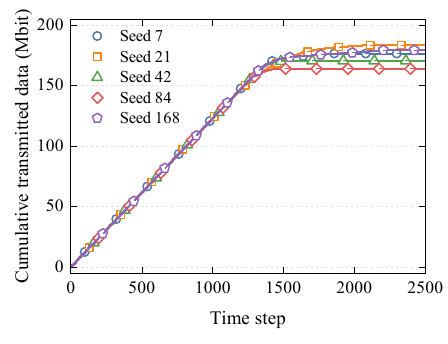}
	\caption{Cumulative transmitted data under HERO-D3QN across five seed settings.}
	\label{fig:seed-throughput}
\end{figure}

The cumulative volumes in Fig.~\ref{fig:seed-throughput} initially followed the same trajectory before diverging across the five seed settings. By time step 999, each run had accumulated $122.88\,\mathrm{Mbit}$ while all 100 nodes remained alive. This common initial contribution accounted for most of the final total in each run, which helps explain the relatively small variation in transmitted volume reported in Fig.~\ref{fig:seed-metric-distribution-b}.
By time step 1494, the curves for seeds 42 and 84 had reached their plateaus. Seed 21 subsequently added $12.32\,\mathrm{Mbit}$, compared with $6.50\,\mathrm{Mbit}$ under seed 168. The larger increase under seed 21 coincided with its larger surviving population during much of the decline in Fig.~\ref{fig:seed-alive}. Seed 168 continued transmitting after seed 21 reached LND, but this final extension added only $1.08\,\mathrm{Mbit}$. The larger accumulation under seed 21 during the decline therefore outweighed the additional transmissions sustained by the last node under seed 168.

\begin{figure}[!tbp]
\centering
\includegraphics[width=0.8\linewidth]{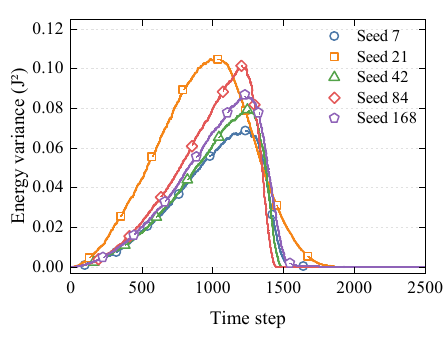}
\caption{Residual energy variance under HERO-D3QN across five seed settings.}
\label{fig:seed-variance}
\end{figure}

\begin{figure}[!tbp]
\centering
\includegraphics[width=0.8\linewidth]{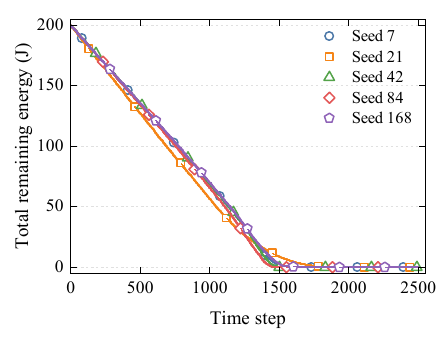}
\caption{Total remaining energy under HERO-D3QN across five seed settings.}
\label{fig:seed-energy}
\end{figure}

Fig.~\ref{fig:seed-variance} shows differences in residual energy dispersion before node failures began. At time step 800, all 100 nodes remained alive in each run, yet seed 21 already had the highest variance. The variance thus revealed differences in energy distribution that were not yet apparent from node survival. Seed 21 subsequently reached the earliest and highest peak, $0.1049\,\mathrm{J}^2$ at time step 1036, whereas seed 7 had the lowest peak, $0.0688\,\mathrm{J}^2$.
After the peaks, the decline in absolute variance also reflected the shrinking energy reserve shown in Fig.~\ref{fig:seed-energy}; variance reached zero at complete energy exhaustion. Seed 21 produced the largest cumulative transmitted volume in Fig.~\ref{fig:seed-throughput}, whereas seed 7 had a lower peak variance but transmitted less. Across these seed settings, lower peak variance did not consistently coincide with a larger cumulative transmitted volume.

The energy trajectories in Fig.~\ref{fig:seed-energy} separated while the curves for survival and cumulative transmitted volume still coincided. At time step 999, each run still had 100 surviving nodes and had accumulated $122.88\,\mathrm{Mbit}$ of transmitted data, yet total remaining energy ranged from $56.6\,\mathrm{J}$ for seed 21 to $70.0\,\mathrm{J}$ for seed 168. The same cumulative volume was therefore associated with different total energy expenditures across seed settings.
The ordering of energy reserves changed as node failures accumulated. At time step 1499, seed 21 retained $8.68\,\mathrm{J}$, compared with $2.82\,\mathrm{J}$ under seed 168 and $1.85\,\mathrm{J}$ under seed 7, while seeds 42 and 84 had exhausted their energy. Relative to seeds 7 and 168, seed 21 had both a larger surviving population in Fig.~\ref{fig:seed-alive} and a higher mean residual energy among its survivors. This combination is consistent with its larger subsequent increase in transmitted volume in Fig.~\ref{fig:seed-throughput}.

\section{Conclusion}\label{sec:conclusion}

This paper presented HERO-D3QN for inter-cluster next-hop selection in WSNs with a static sink. Given the cluster structure at each time step, the method combines an enhanced state representation of relay energy, progress toward the sink, potential forwarding load, and hotspot risk with a multi-objective reward. Its action selector screens candidate relays by direction and forwarding cost, then applies soft penalties to their estimated action values. Relays with low residual energy remain eligible when they satisfy the other selection criteria.

In the comparison using the same node locations, HERO-D3QN reached FND at time step 1269, approximately 2.02 times the value for D3QN, with a 55.5\% lower peak residual energy variance. This extension of full node survival was accompanied by an earlier LND and a final transmitted volume of 170.52 Mbit, compared with 175.07 Mbit for D3QN. Among the tested variants, the full method achieved the latest FND and HND and the largest transmitted volume, while the basic-state and direct intra-cluster transmission variants had lower peak variances. Across five seed settings in which node locations and the random processes used in training and evaluation varied jointly, the CVs of HND and transmitted volume were 2.75\% and 3.88\%, respectively, with greater relative variation in LND and energy variance. These results identify the main benefit of HERO-D3QN as delaying early node failure while limiting energy dispersion, with distinct outcomes for the persistence of the last surviving nodes and cumulative transmitted volume.

The present evaluation assumes static node locations and a fixed sink. Future work will refine replay using consecutive routing transitions, assess end-to-end data delivery, compare component effects across seeds, and examine alternative clustering mechanisms and dynamic link conditions.

\backmatter

\bmhead{Supplementary information}

The simulation code and configurations that support the findings of this study will be made publicly available as a GitHub repository upon acceptance.

\bmhead{Acknowledgements}

This work was supported in part by the Natural Science Foundation of Liaoning Province under Grant 2024MSLH359 and Grant 2025MSLH547; in part by the Scientific Research Fund of Liaoning Provincial Education Department under Grant LJ212410142021 and Grant LJ212510142045; in part by the University-Industry Collaborative Education Program of the Ministry of Education under Grant 231106665030834 and Grant 231106665072727.

\section*{Declarations}

The authors have no relevant financial or non-financial interests to disclose.


%\bibliography{references}% common bib file


\end{document}